\documentclass[aps,prd,twocolumn,superscriptaddress]{revtex4-1}

\usepackage{xspace}
\usepackage{xcolor}
\usepackage{graphicx}
\usepackage{amssymb}
\usepackage{amsmath}
\usepackage{bm}
\usepackage{epsfig}
\usepackage[section]{placeins}
\usepackage[export]{adjustbox}

\newcommand{\bea}{\begin{eqnarray}}
\newcommand{\eea}{\end{eqnarray}}

\def\be{\begin{eqnarray}}
\def\ee{\end{eqnarray}}

\def\pslash{\not{\hbox{\kern-2pt p}}}
\newcommand{\met}{\ensuremath{{\not\mathrel{\vec{E}}}_T}\xspace}

\definecolor{mygreen}{rgb}{0.12, 0.3, 0.17}
\definecolor{myred}{rgb}{0.45, 0.0, 0.09}
\definecolor{myblue}{rgb}{0.1, 0.1, 0.44}
\newcommand{\Vframe}{\ensuremath{\mathrm{\textcolor{myblue}{\textbf{V}}}}\xspace}
\newcommand{\ISRframe}{\ensuremath{\mathrm{\textcolor{myblue}{\textbf{ISR}}}}\xspace}
\newcommand{\CMframe}{\ensuremath{\mathrm{\textcolor{myred}{\textbf{CM}}}}\xspace}
\newcommand{\Sframe}{\ensuremath{\mathrm{\textcolor{myred}{\textbf{S}}}}\xspace}
\newcommand{\Iframe}{\ensuremath{\mathrm{\textcolor{mygreen}{\textbf{I}}}}\xspace}

\newcommand{\RISR}{\ensuremath{R_{\scriptsize{\mathrm{ISR}}}}\xspace}
\newcommand{\PTISR}{\ensuremath{p_{\scriptsize{\ISRframe},T}^{~\scriptsize{\CMframe}}}\xspace}
\newcommand{\MS}{\ensuremath{M_{T}^{\scriptsize{\Sframe}}}\xspace}
\newcommand{\NjV}{\ensuremath{N_{\mathrm{jet}}^{\scriptsize{\Vframe}}}\xspace}

\newcommand{\dphiISRI}{\ensuremath{\Delta \phi_{\scriptsize{\ISRframe,\Iframe}}}\xspace}

\begin{document}


\title{Sparticles in Motion - getting to the line in compressed scenarios with the Recursive Jigsaw Reconstruction}

\author{Paul Jackson}
\email{p.jackson@adelaide.edu.au}
\affiliation{University of Adelaide, Department of Physics, Adelaide, SA 5005, Australia}
\affiliation{ARC Centre of Excellence for Particle Physics at the Tera-scale, University of Adelaide}
\author{Christopher Rogan}
\email{crogan@physics.harvard.edu}
\affiliation{Harvard University, Department of Physics, 17 Oxford Street, Cambridge, MA 02138}
\author{Marco Santoni}
\email{marco.santoni@adelaide.edu.au}
\affiliation{University of Adelaide, Department of Physics, Adelaide, SA 5005, Australia}
\affiliation{ARC Centre of Excellence for Particle Physics at the Tera-scale, University of Adelaide}

\date{\today}

\begin{abstract}
The observation of light super-partners from a supersymmetric extension to the Standard Model is an intensely sought-after experimental outcome, providing an explanation for the stabilization of the electroweak scale and indicating the existence of new particles which could be consistent with dark matter phenomenology. 
For compressed scenarios, where sparticle spectra mass-splittings are small and decay products carry low momenta,
dedicated techniques are required in {\it all} searches for supersymmetry. In this paper we suggest an approach for these analyses based on the concept of {\it Recursive Jigsaw Reconstruction}, decomposing each event into a basis of complementary observables, for cases where strong initial state radiation has sufficient transverse momentum to elicit the recoil of any final state sparticles. We introduce a collection of kinematic observables which can be used to probe compressed scenarios, in particular exploiting the correlation between missing momentum and that of radiative jets.  As an example, we study squark and gluino production, focusing on mass-splittings between parent super-particles and their lightest decay products between 25 and $200$~GeV, in hadronic final states where there is an ambiguity in the provenance of reconstructed jets. 
\end{abstract}

\pacs{Pacs numbers}

\maketitle

\section{Introduction\label{sec:intro}}
Armed with large datasets of high energy collisions, experimentalists have a multitude of choices of how to search for evidence of physics that cannot be explained by the Standard Model (SM) theory. 
Canonical solutions to some of the shortcomings of the SM suggest a conserved $\mathbb{Z}_2$ symmetry of some type, which has important phenomenological consequences for observing evidence of new particles. 
A key implication is that there must be an even number of new particles in each interaction, further implying that they are produced in pairs at colliders. By this same restriction, the lightest new particle, unable to decay to SM particles, would be stable. In Supersymmetry~\cite{ref:Golfand1971iw, ref:Neveu1971rx, ref:Ramond1971gb, ref:Volkov1973ix, ref:Wess1974tw} (SUSY), the lightest weak-scale SUSY particle (LSP) could be a candidate for dark matter. If produced at colliders, LSPs will leave the detectors without interacting. Their presence can be inferred through measurement of missing momentum, particularly in the plane transverse to the beam-line at hadron colliders. In practice, extracting information about the properties of the LSP, or the multiplicity of weakly-interacting particles in each event, presents a challenge to reconstruction techniques. The difficulty is exacerbated in situations where there is a small mass difference between the pair-produced parent sparticles and their lightest sparticle decay products. Such a decay chain scenario is colloquially referred to as being {\it compressed}. In this case, the impact of handles typically exploited to separate signal 
processes from their SM backgrounds, such as large object transverse momentum and missing transverse momentum, is compromised, as the majority of the energy from sparticle decays escapes detection in the mass of the LSPs.

In this paper, we introduce a set of experimental observables which can be used to increase sensitivity to these compressed scenarios at collider experiments. As described in Sec.~\ref{sec:motion}, they are designed for the analysis of events where one or more strong initial state radiation (ISR) jets are present, causing the system of initially produced sparticles to recoil in the opposite direction. Using a simplified view of the event, a basis of kinematic observables is derived to exploit kinematic correlations induced by the presence of massive LSPs, calculated from the missing transverse momentum and particle four-vectors reconstructed by detectors. While this approach is straight-forward when SM decay products of sparticles in these events are readily identifiable, decays to jets result in a potential ambiguity; the provenance of each jet, ISR or sparticle, is unknown. In Sec.~\ref{sec:jets} we describe a strategy for overcoming this difficulty, using as an example squark and gluino pair-production with mass splittings between the parent sparticle and LSP of $25 - 200$ GeV. 

\section{Sparticles in motion\label{sec:motion}}

Motivated by R-parity and phenomenologically similar new physics models with a $\mathbb{Z}_2$ symmetry, we consider cases where initial sparticle parents ($\tilde{P}$) are pair-produced, and each decay to a system of reconstructable SM particles and one, or more, weakly-interacting ones. In the following discussion, we assume for simplicity that the masses of these parents are identical, as are the masses their weakly interacting daughters ($\tilde{\chi}^{0}$). An experimental search for instances of these events can be difficult if the mass-splitting between these sparticle states, $m_{\tilde{P}}-m_{\tilde{\chi}^{0}}$, is small, as the momenta of each parent sparticle's decay products (both visible and invisible) will not receive a large amount of momentum in their production. If the mass-splitting scale in sparticle production is to that of SM background processes then disentangling the two is challenging. 

In this case, it is not the mass-splitting scale which is distinctive from backgrounds, but rather, the potentially large absolute mass-scale of weakly-interacting particles in these events. While we cannot measure these masses from only the measurement of missing transverse momentum (\met), as it only represents the sum momentum of escaping particles, we can gain indirect sensitivity by observing their reaction to a probing force. The laboratory of a hadron collider naturally provides such a probe: strong initial state radiation from interacting partons can provide large momentum to the sparticles produced in these reactions, in turn endowing their decay products with this momentum. In the limit where the LSPs receive no momentum from their parents' decays, the \met results solely from the recoil against ISR, and the following approximation holds:
\begin{equation}
\met \sim -\vec{p}_{T}^{\rm  ~ISR} \times \frac{m_{\tilde{\chi}}}{m_{\tilde{P}}}~,
\label{eqn:met}
\end{equation} 
where $\vec{p}_{T}^{\rm  ~ISR}$ is the total ISR system transverse momentum.

Recent studies of searches for compressed SUSY signals in the literature have suggested exploiting this feature. In these analyses, a kinematic selection is used to isolate events where a single, hard ISR jet recoils approximately opposite \met in the event transverse plane. One can then use various reconstructed proxies of the quantity $|\met|/|\vec{p}_{T}^{\rm  ~ISR}|$, such as $|\met|/p_{T}^{~\mathrm{lead~jet}}$ or $|\met|/\sqrt{H_{T}}$, as observables sensitive to the presence of massive LSPs~\cite{ref:compressed_stop-1,ref:compressed_stop-2}. Alternatively, using assumed knowledge of the sparticle mass-splittings, one can attempt to sort non-ISR jets from radiative ones using, for example, the sum of jet energies in each class and multiplicities as discriminating observables~\cite{ref:compressedgluinos}. While these approaches all benefit from the above feature, they are limited to the sub-set of events where the momentum of the ISR system is carried predominantly by a single jet. For less restrictive event selections, the suggested observables become progressively less accurate estimators of $|\met|/|\vec{p}_{T}^{\rm  ~ISR}|$ and, correspondingly, less sensitive to the kinematic correlation between radiated jets and missing momentum. 

\begin{figure}[tbh!]
\centering 
\includegraphics[width=.4\textwidth]{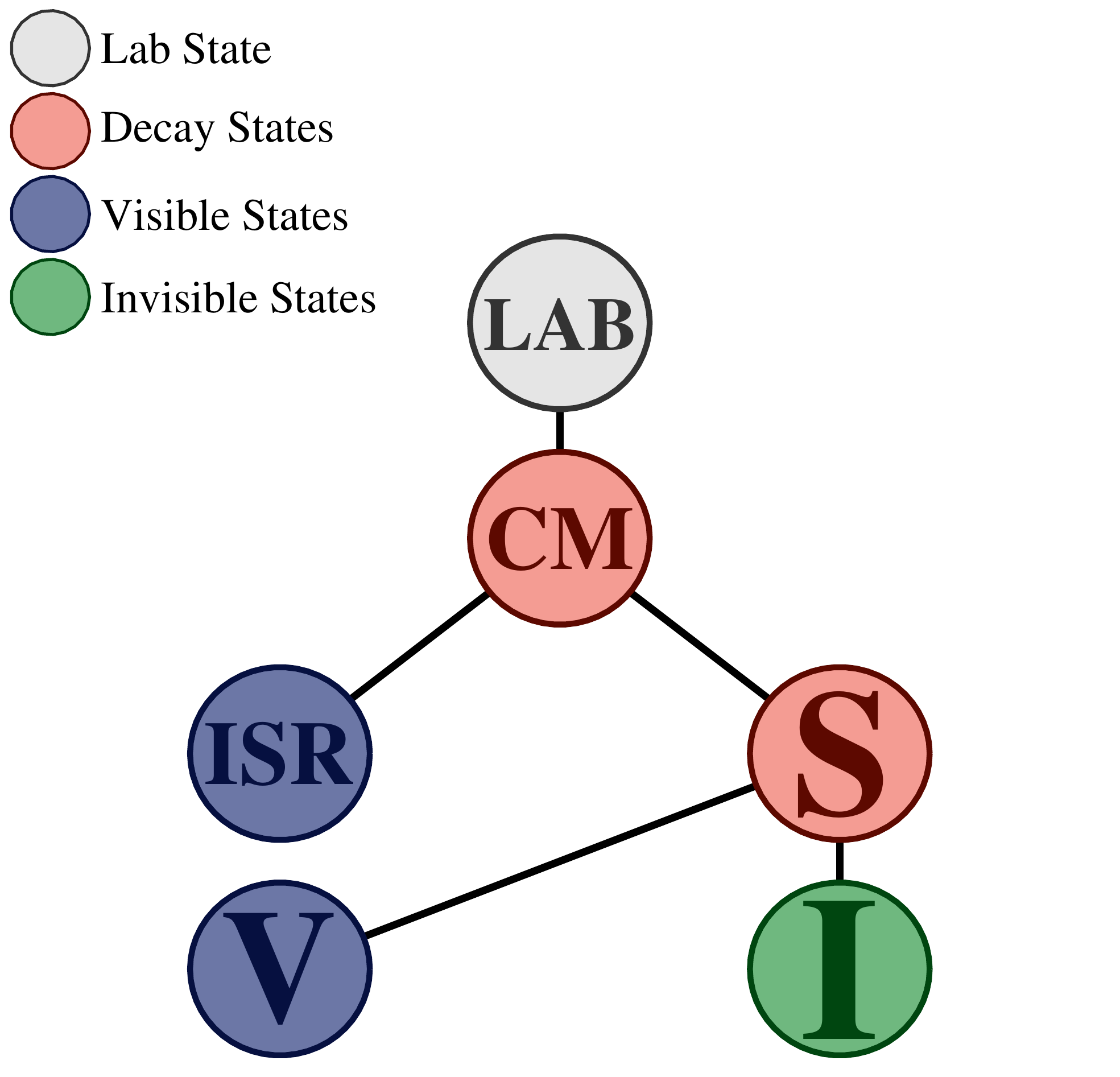}
\caption{\label{fig:decayTree} A simplified decay tree diagram for analyzing compressed signal topologies in events with an ISR system.}
\end{figure}

We propose a different approach to an ISR-assisted search for compressed signals, both generalizing to cases where momentum can be shared democratically among many radiated jets and attempting to more accurately reconstruct the quantity $|\met|/|\vec{p}_{T}^{\rm  ~ISR}|$. Using the technique of {\it Recursive Jigsaw Reconstruction}~\cite{ref:basis}, a ``decay tree'' is imposed on the analysis of each event, chosen to capture the kinematic features specific to the signal topology under study. The decay tree both specifies the systems of relevant reconstructed objects and the reference frames corresponding to each intermediate combination of them. The analysis of each event proceeds by assigning reconstructed objects to their appropriate places in the decay tree, determining the relative velocities relating each reference frame, and calculating kinematic observables from the resulting event abstraction. The simplified decay tree for generic compressed scenarios is shown in Figure~\ref{fig:decayTree}.

In this decay tree, each reconstructed object hypothesized to come from the decay of sparticles in the event is assigned to the ``\Vframe'' system, while those identified as initial state radiation are associated with ``\ISRframe''. With the missing momentum reconstructed in each event interpreted as the system ``\Iframe'', the total sparticle system (``\Sframe'') and center-of-mass system of the whole reaction (``\CMframe'') are defined as the sum of their constituents. With the four-vectors of each element of the decay tree specified, an estimator of the quantity $|\met|/|\vec{p}_{T}^{\rm  ~ISR}|$, \RISR, is calculated as:
\begin{equation}
\RISR \equiv \frac{|\vec{p}_{\scriptsize{\Iframe},T}^{~\scriptsize{\CMframe}}\cdot \hat{p}_{\scriptsize{\ISRframe},T}^{~{\mathrm{\scriptsize{\CMframe}}}}|}{|\vec{p}_{\scriptsize{\ISRframe},T}^{~{\mathrm{\scriptsize{\CMframe}}}}|}~,
\label{eqn:RISR}
\end{equation} 
where subscripts indicate the system and superscripts the reference frame the momentum is evaluated in. As the concept of ``transverse'' is a frame-dependent construction in the laboratory frame, we employ the convention where the boost relating a specific reference frame to the laboratory is decomposed into a component parallel to the beam-line and a subsequent transverse portion. The transverse plane in a reference frame is then defined as that perpendicular to longitudinal velocity of the transformation.

\begin{figure}[tbh!]
\centering 
\hspace{-.087\textwidth}
\includegraphics[width=.42\textwidth]{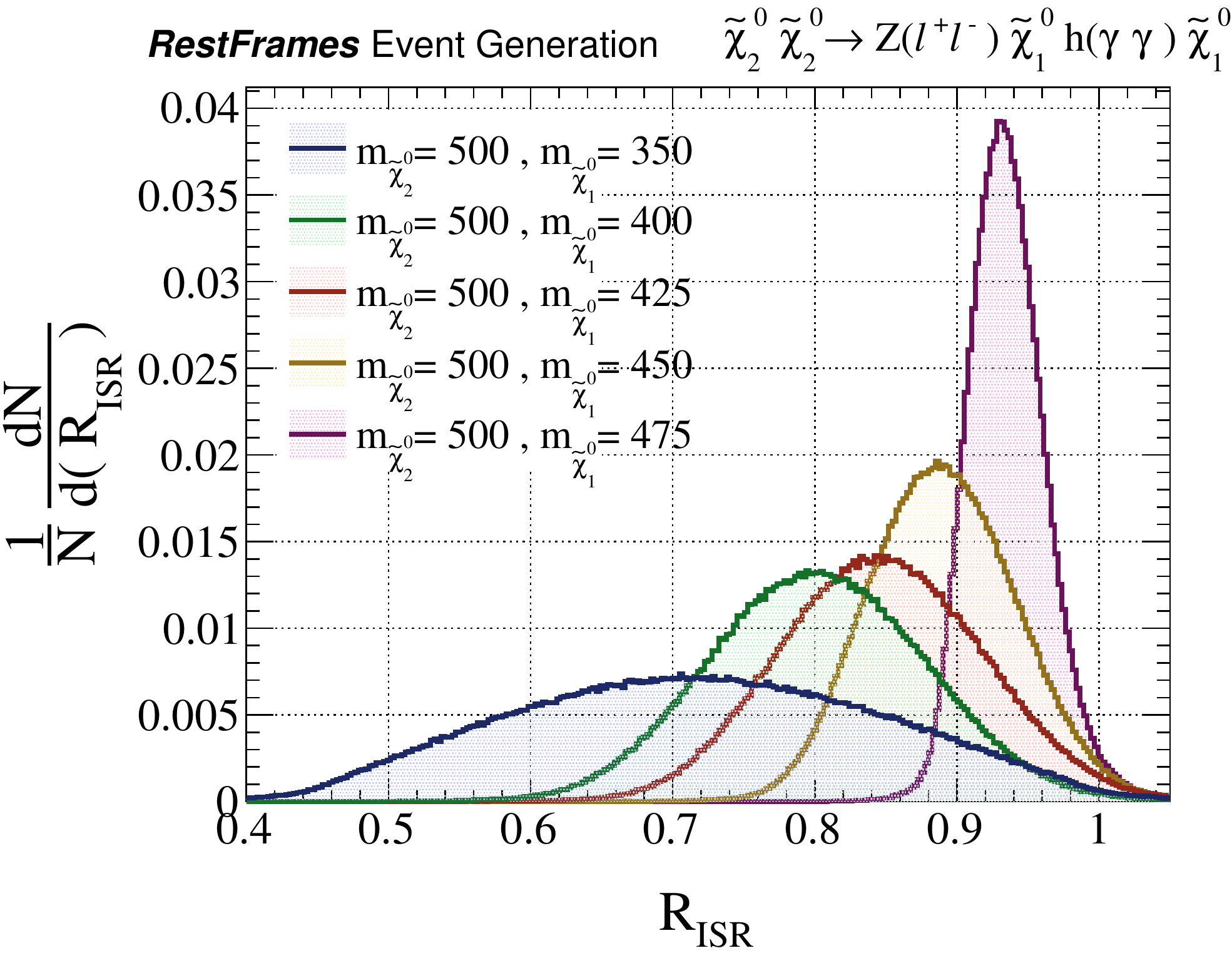}
\includegraphics[width=.485\textwidth]{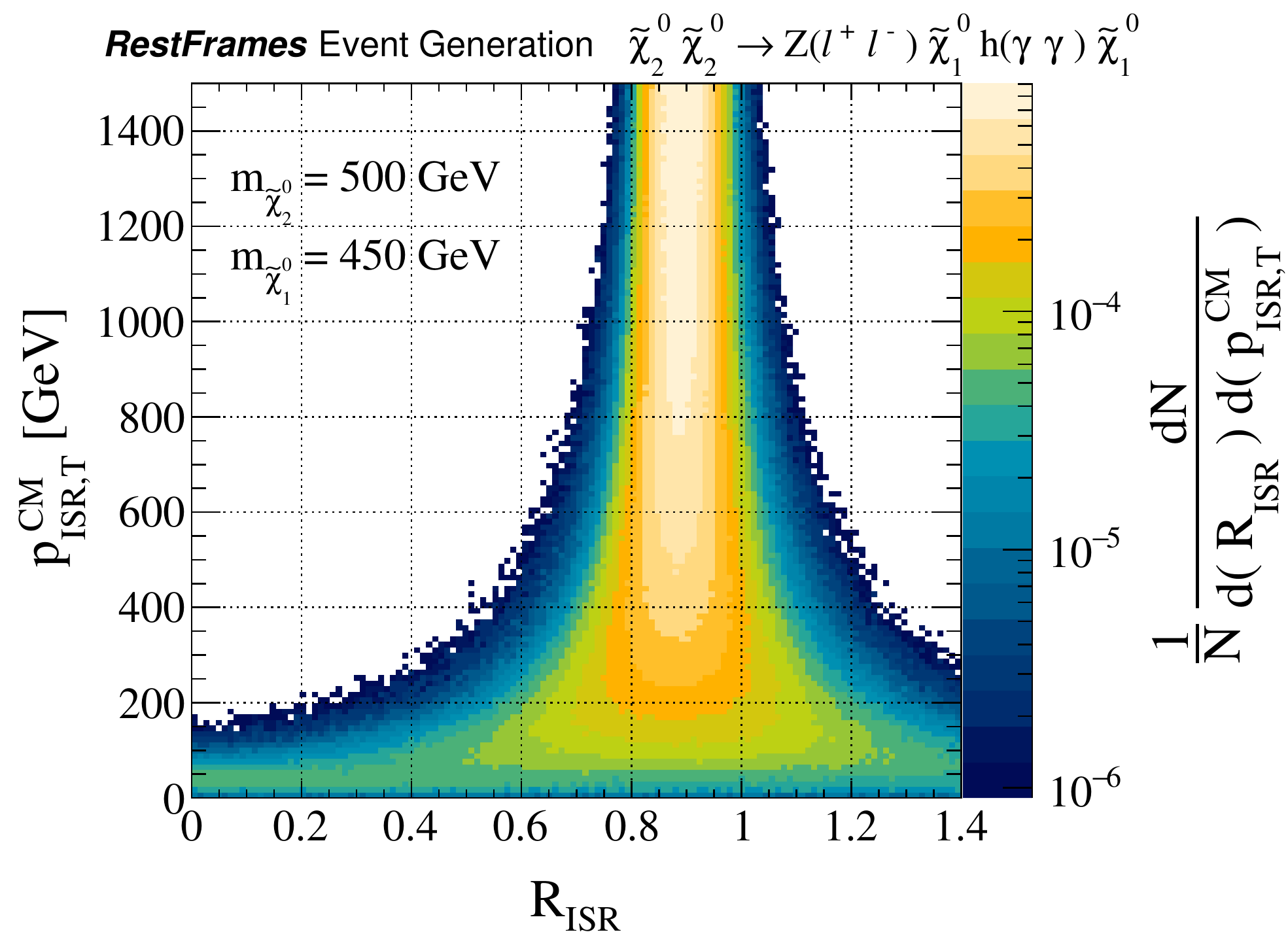}
\caption{\label{fig:RISR_EWKino} The distribution of \RISR for the production and decay of $\tilde{\chi}^{0}_{2}\tilde{\chi}^{0}_{2}\rightarrow Z( \ell^{+}\ell^{-})\tilde{\chi}^{0}_{1}~h( \gamma\gamma)\tilde{\chi}^{0}_{1}$ at the 13 TeV LHC. (Top) \RISR for $m_{\tilde{\chi}^{0}_{2}} = 500$ GeV and varying $m_{\tilde{\chi}^{0}_{1}}$. (Bottom) \RISR as a function of \PTISR for  $m_{\tilde{\chi}^{0}_{2}} = 500$ GeV, $m_{\tilde{\chi}^{0}_{1}} = 450$ GeV. Simulated events are generated and analyzed using the \texttt{RestFrames} software package~\cite{ref:restframes}.}
\end{figure}

In order to elucidate the behavior of \RISR, we consider the example of neutralino ($\tilde{\chi}^{0}_{2}$) pair-production at a hadron collider with decays $\tilde{\chi}^{0}_{2}\tilde{\chi}^{0}_{2}\rightarrow Z( \ell^{+}\ell^{-})\tilde{\chi}^{0}_{1}~h( \gamma\gamma)\tilde{\chi}^{0}_{1}$. Two leptons and two photons are required to be reconstructed in each event and are assigned to the \Vframe system, while additional reconstructed jets are associated with \ISRframe. \met reconstructed in each event is interpreted as the transverse momentum of the \Iframe system in the laboratory frame, with zero mass and rapidity set equal to that of the \Vframe system. The \RISR distributions for simulated events, for varying sparticle masses, are shown in Figure~\ref{fig:RISR_EWKino}.

We observe that the \RISR distribution for these events scales with $m_{\tilde{\chi}^{0}_{1}}/m_{\tilde{\chi}^{0}_{2}}$, as expected from Eq.~\ref{eqn:met}, with increasingly fine resolution for progressively smaller mass-splittings between the two sparticle states. Similarly, the resolution of the kinematic feature improves for larger values of \PTISR. This behavior can be understood from a more careful examination of the approximate relation described in Eq.~\ref{eqn:met}. In the limit that the momenta of the sparticle daughter in its parent's rest frame, $p_{\tilde{\chi}^{0}}^{\tilde{P}}$, is small relative to $m_{\tilde{P}}$, \RISR corresponds to:
\begin{eqnarray}
\RISR \sim \frac{|\met \cdot \hat{p}_{T}^{\rm ~ISR}|}{p_{T}^{\rm ~ISR}} \sim \quad  \quad \quad \quad \quad \quad \quad \quad \quad \quad \quad \quad \quad~~ \\
 \frac{m_{\tilde{\chi}^0}}{m_{\tilde{P}}} \left[ 1 + \mathcal{O}\left( \frac{p_{\tilde{\chi}^0}^{\tilde{P}}}{2m_{\tilde{P}}}\right)\left(\frac{\sqrt{(p_{T}^{\rm~ISR})^{2}+m_{\tilde{P}\tilde{P}}^{2}}}{p_{T}^{\rm~ISR}}\right) \mathrm{sin}~\Omega  \right]~, \nonumber
\label{eqn:RISR}
\end{eqnarray} 
where $\mathrm{sin}~\Omega$ represents order one dot products between the velocities relating the laboratory frame, the $\tilde{P}\tilde{P}$ rest frame, and $\tilde{P}$ rest frames and which, in the absence of non-trivial spin correlations or efficiency dependence from decay product reconstruction and selection, is zero on average. \RISR scales with the ratio $m_{\tilde{\chi}^{0}}/m_{\tilde{P}}$, with resolution of the order $p_{\tilde{\chi}^0}^{\tilde{P}}/2m_{\tilde{P}}$ in the limit $p_{T}^{\rm~ISR} \gg m_{\tilde{P}\tilde{P}}$.

Hence, the observable \RISR is an excellent proxy to the quantity $|\met|/|\vec{p}_{T}^{\rm  ~ISR}|$, sensitive to the presence of massive LSPs in the event, with accuracy improving with increasing compression between sparticle masses. To observe this behavior, the sparticles must be put in motion by a probe with measurable momenta, in this case an ISR system of strong radiation, with harder probes yielding better resolution of the $m_{\tilde{\chi}^0}/m_{\tilde{P}}$ ratio. By construction, this approach does not require that the ISR system's momenta is contained predominantly in a single jet. Of course, generalizing to final states with many ISR jets is not without its challenges. In the above example, there is no ambiguity in which reconstructed objects should be assigned to the sparticle and \ISRframe systems, respectively.  In the following section, we describe a strategy for analyzing events where there are combinatoric ambiguities in the assignment of reconstructed objects to our decay tree. We explain how to calculate \RISR in these cases, along with a set of complementary observables designed for analyzing compressed event topologies. 

\section{Compressed squark and gluino production with hadronic decays \label{sec:jets}}

When the visible decay products resulting from compressed sparticle production are identifiable by their type, the interpretation according to the decay tree in Figure~\ref{fig:decayTree} is straight-forward; reconstructed objects expected to come from sparticles are assigned to the \Vframe system, while jets correspond to \ISRframe. Unfortunately, all signal topologies of interest may not be this simple, as sparticles can also decay directly or indirectly to SM partons, resulting in reconstructed jets indiscernible from ISR. In the context of Recursive Jigsaw Reconstruction, such a combinatoric ambiguity is resolved through the application of a ``{\it jigsaw rule}'', an interchangeable algorithm for assigning indistinguishable final-state objects to different sub-systems of a decay tree. 

Motivated by our qualitative understanding of events with an ISR system recoiling against a compressed sparticle system, we apply a jigsaw rule which attempts to group objects together which are nearby in phase-space, effectively minimizing the reconstructed masses of the \Sframe and \ISRframe systems. Specifically, we employ an exclusively transverse view of each event, ignoring the longitudinal momenta of all the reconstructed objects. The mass of the \Iframe~system is approximated to be zero, with transverse momenta in the laboratory frame set equal to \met. The total (transverse) mass of the \CMframe system can then be expressed as
\begin{equation}
M_{\mathrm{\scriptsize{\CMframe}}} = \sqrt{M_{\mathrm{\scriptsize{\ISRframe}}}^2+\left(p_{\mathrm{\scriptsize{\ISRframe}}}^{~\mathrm{\scriptsize{\CMframe}}}\right)^2} + \sqrt{M_{\mathrm{\scriptsize{\Sframe}}}^2+\left(p_{\mathrm{\scriptsize{\Sframe}}}^{~\mathrm{\scriptsize{\CMframe}}}\right)^2}~,
\label{eq:MinMasses}
\end{equation}
with $p_{\mathrm{\scriptsize{\Sframe}}}^{~\mathrm{\scriptsize{\CMframe}}}$ and $p_{\mathrm{\scriptsize{\ISRframe}}}^{~\mathrm{\scriptsize{\CMframe}}}$ the (equal) magnitudes of the momentum of the \Sframe and \ISRframe systems, respectively, evaluated in the \CMframe frame, and dependent on our choice of combinatoric assignment of objects. As $M_{\mathrm{\scriptsize{\CMframe}}}$ does not depend on this assignment, we effectively minimize $M_{\mathrm{\scriptsize{\Sframe}}}$ and $M_{\mathrm{\scriptsize{\ISRframe}}}$ simultaneously by maximizing $p_{\ISRframe/\Sframe}^{~\CMframe}$ over each potential partitioning of indistinguishable objects into either the \Vframe or \ISRframe systems. Qualitatively, this is similar to treating the \met as another reconstructed object and performing an exclusive jet-clustering, using the transverse mass as a distance metric. 

While this approach does not distinguish between ISR and sparticle jets with perfect efficiency, it does provide a unique, deterministic assignment of objects to our compressed decay tree. Furthermore, in addition to \RISR, we can extract an entire collection of complementary observables from this event interpretation, chosen to further discriminate between putative compressed sparticle signals and SM backgrounds. These observables include:
\begin{itemize}
\item \PTISR: the magnitude of the the vector-sum transverse momentum of all \ISRframe associated jets, evaluated in the \CMframe~frame
\item \MS: the transverse mass of the $\Sframe$ $(\Vframe + \Iframe)$ system. 
\item \NjV: the number of jets assigned to the \Vframe system (i.e. not associated with the \ISRframe system)
\item \dphiISRI: the opening angle between the \ISRframe~system and the \Iframe system, evaluated in the \CMframe~frame.
\end{itemize}

To demonstrate the efficacy of a search analysis based on this set of observables, we examine perhaps the most difficult analysis scenario: sparticles expected to decay {\it exclusively} to jets and weakly interacting particles.  Specifically, we study the phenomenology of pair-produced squarks and gluinos decaying to quarks and LSPs ($\tilde{q}\tilde{q} \rightarrow (q\tilde{\chi}^0)(q\tilde{\chi}^0)$ and $\tilde{g}\tilde{g} \rightarrow (qq\tilde{\chi}^0)(qq\tilde{\chi}^0)$). Simulated Monte Carlo (MC) samples of SM backgrounds and SUSY signals are used to construct the expected distributions of these observables for various processes. We utilize background samples from elsewhere~\cite{ref:snowmass-1}.  
For these samples, event generation is performed with Madgraph~5~\cite{ref:madgraph}, along with parton shower and hadronization with Pythia~6~\cite{ref:pythia}. This is followed by a detailed detector simulation and description of pile-up with Delphes~3~\cite{ref:delphes}. A detector parameterization is used which incorporates the performance of the existing ATLAS~\cite{ref:atlas} and CMS~\cite{ref:cms} detectors. Each of the SM processes which are expected to constitute the largest backgrounds are considered. The simulation procedure involves generation of events at leading order in bins of the scalar sum of the recoil jet $p_T$ ($H_T$), with jet-parton matching and corrections for next-to-leading order (NLO) contributions. Further details can be found elsewhere~\cite{ref:snowmass-2}. 

Similarly, squark and gluino signal samples are produced, mimicking the procedure employed to create background samples. The gluinos are considered to decay via $\tilde{g}\rightarrow qq\tilde{\chi}_{1}^{0}$ and the squarks via $\tilde{q}\rightarrow q\tilde{\chi}_{1}^{0}$, in what are akin to simplified models, where the branching fractions are assumed to be 100$\%$ and the masses of other non-contributing super-partners are effectively decoupled. We simulate samples with squark masses between 400 and 1000~GeV, and gluino masses from 600 to 1400~GeV. The $\tilde{\chi}_{1}^{0}$ mass is set to be either 25,~50,~100, or 200~GeV below the parent sparticle mass, covering a dynamic range of compressed scenarios.
To study the potential impact of this approach on analyses being performed at the LHC experiments, we normalize all background and signal samples to an integrated luminosity of 100 fb$^{-1}$, such that the estimated sensitivities shown herein should be accessible during run~2 of the LHC.

\begin{figure}[thb!]
\centering 
\includegraphics[width=.46\textwidth]{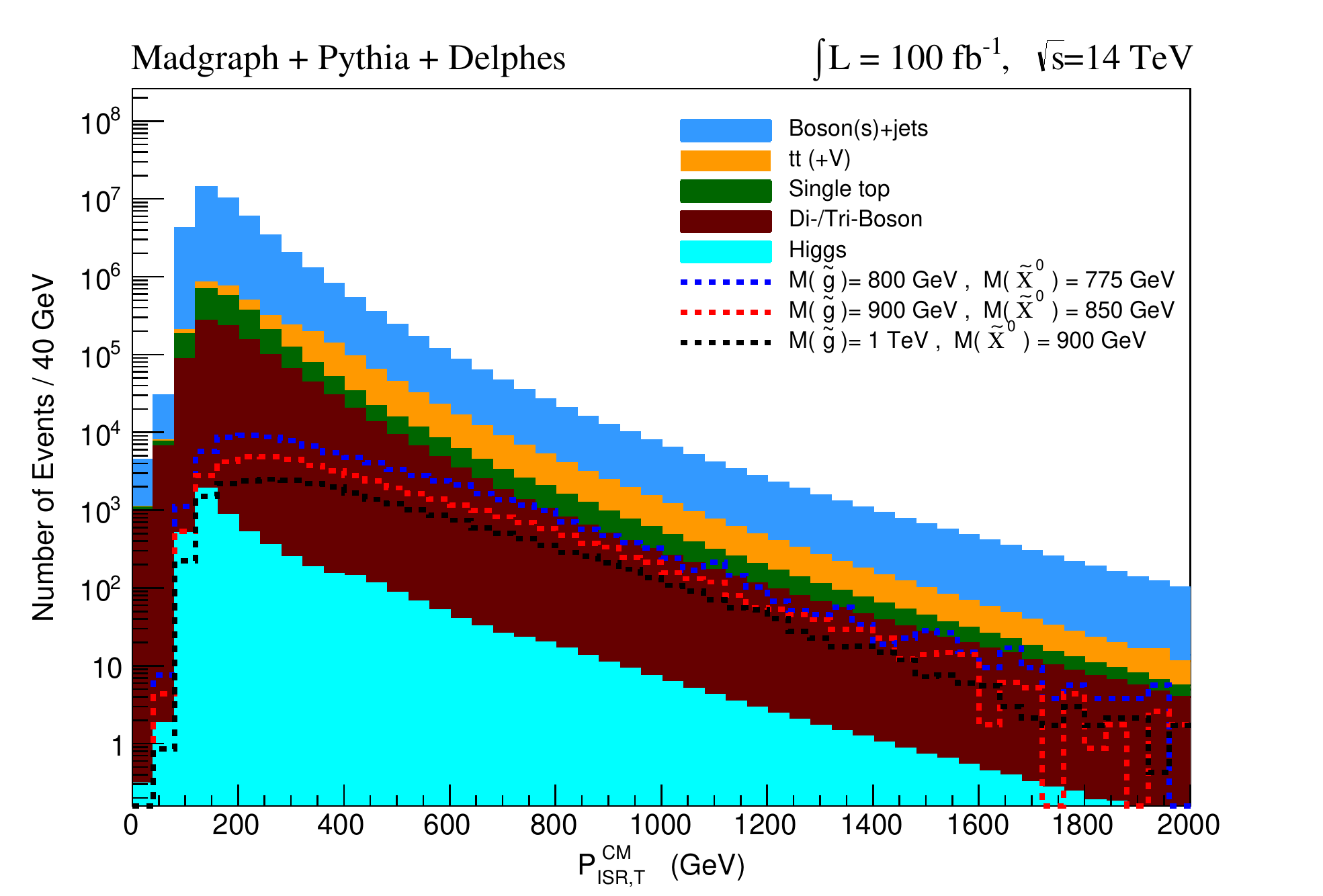}
\caption{\label{fig:pTISR} Distribution of \PTISR for events that have passed the preselection requirements in Table~\ref{Tab:SignalRegions}. A high $p_{T}$ jet system, identified with ISR, is required to build the decay tree.}
\end{figure}

\begin{figure}[htb!]
\vspace{0.2cm}
\centering 
\includegraphics[width=.246\textwidth]{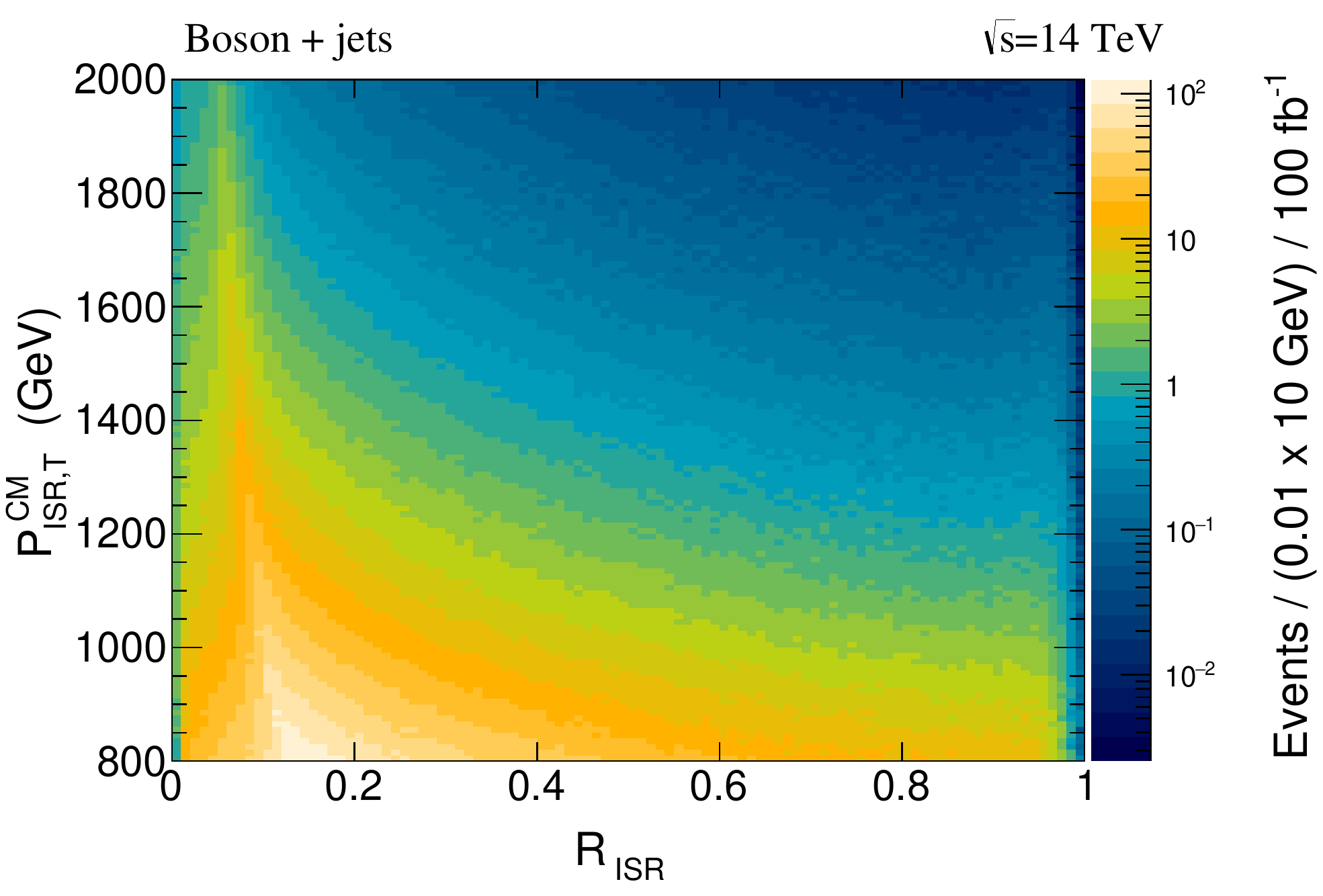}
\includegraphics[width=.246\textwidth]{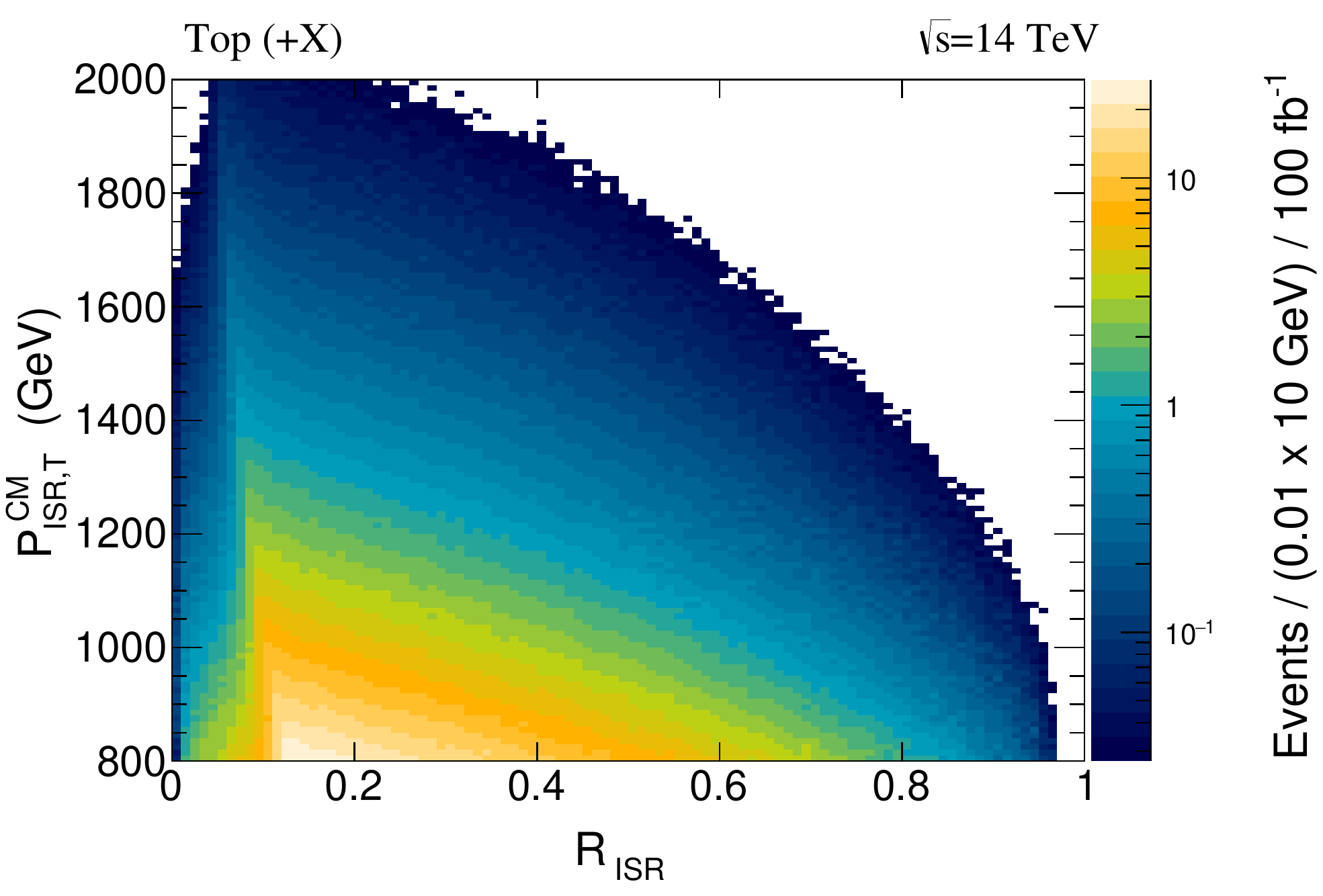}
\includegraphics[width=.246\textwidth]{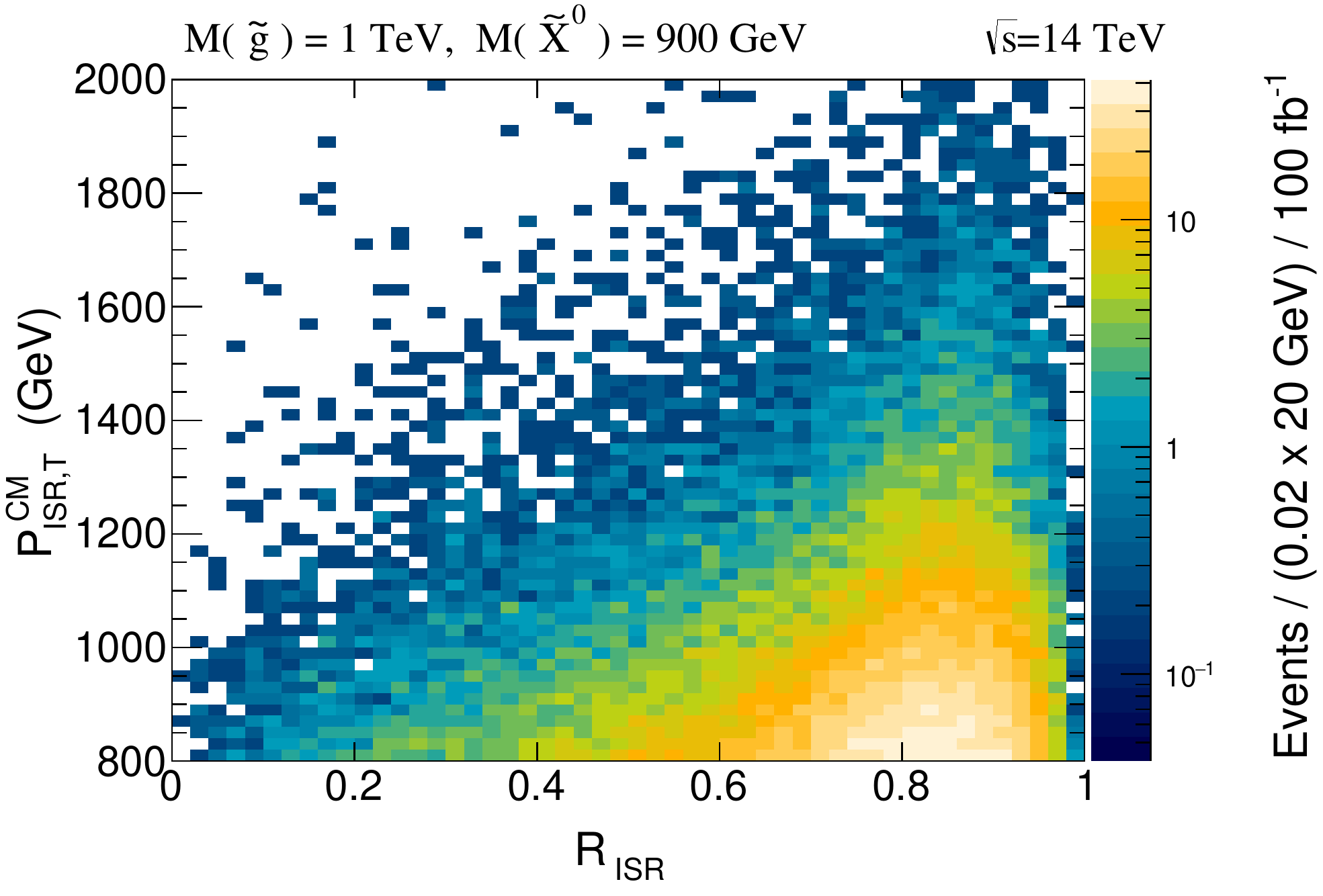}
\includegraphics[width=.246\textwidth]{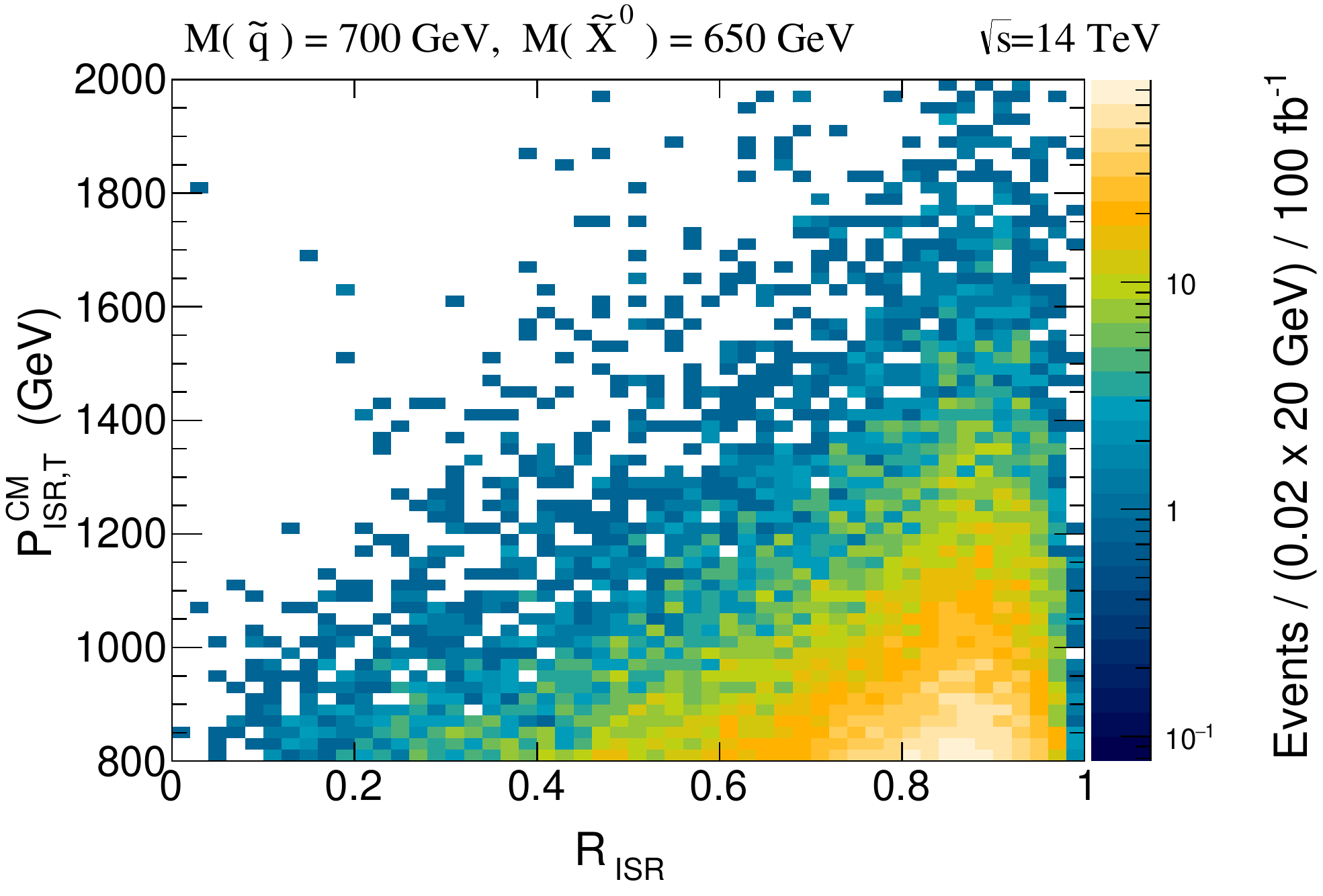}
\caption{\label{fig:RISR_vs_pTISR}  
Distribution of the \PTISR as a function of \RISR  for (from left to right) boson+jets and top+X backgrounds, gluino and squark pair-production signal samples.}
\end{figure}

The distributions of the \PTISR for SM backgrounds and signal MC processes, with different parent sparticle masses and mass-splittings, are shown in Figure~\ref{fig:pTISR}. We observe that, prior to the application of any other selection criteria, prospective signals have small expected event yields relative to SM backgrounds for lower values of \PTISR. However, as the slope of the \PTISR distribution is less severe for these signals, the signal-to-background ratio becomes more favorable with increasing values. In the following, we consider only those events with $\PTISR \ge 800$ GeV, not only taking advantage of the moderate discrimination provided by this observable, but also benefitting from the effect that this requirement has on other, correlated, variables. The most striking example of this complementarity can be seen in the two-dimensional distributions of \PTISR and \RISR, shown in Figure~\ref{fig:RISR_vs_pTISR} for signal and backgrounds. Analogous to Figure~\ref{fig:RISR_EWKino}, increasing \PTISR results in a narrowing of the \RISR distribution for compressed signals, while \PTISR and \RISR are strongly anti-correlated for backgrounds. Hence, progressively stricter \PTISR requirements yield improved \RISR discrimination, with the optimal selection for the latter depending on the signal characteristics, in particular the ratio $m_{\tilde{\chi}^0}/m_{\tilde{P}}$.  
\onecolumngrid

\twocolumngrid

\begin{figure}[htb!]
\centering 
\includegraphics[width=.46\textwidth]{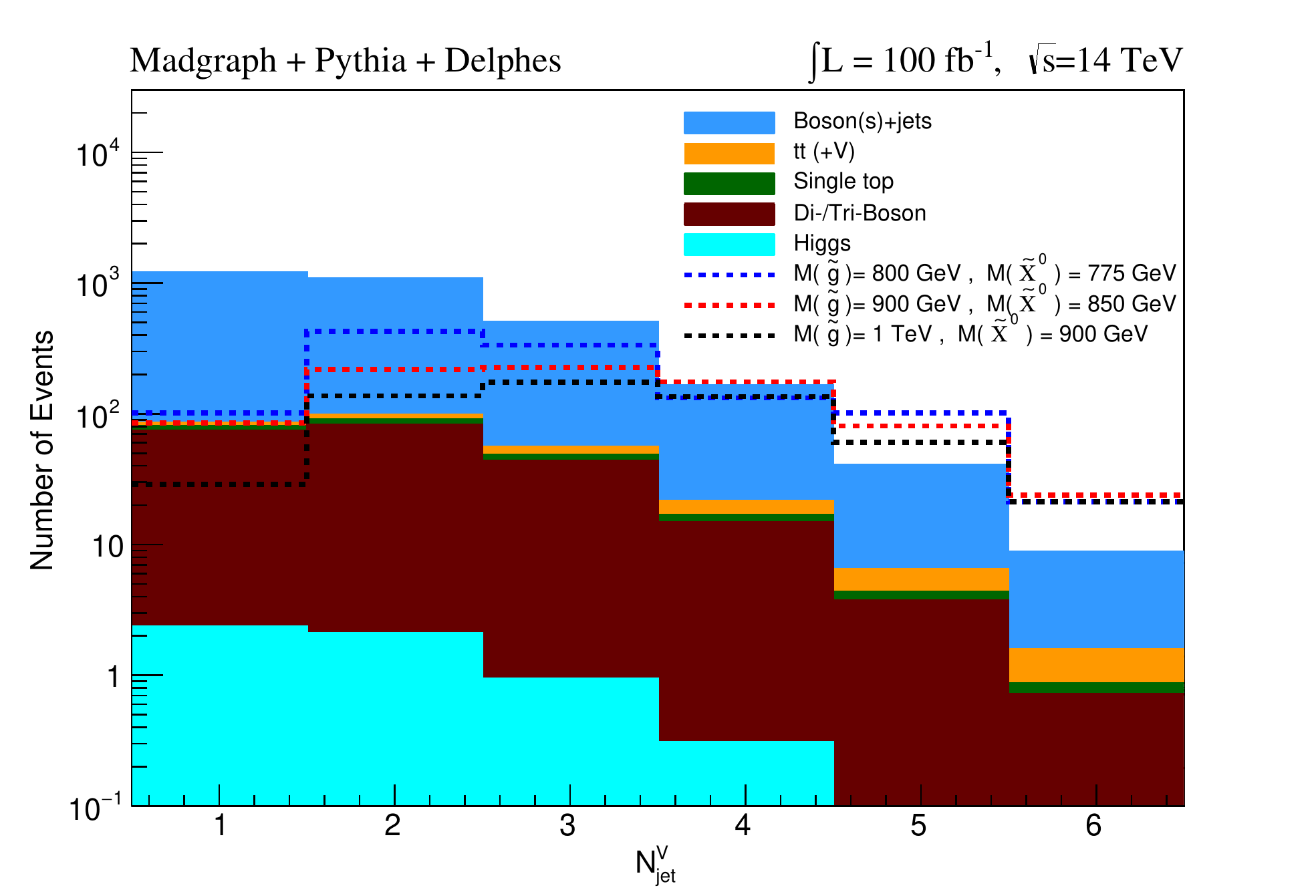}
\caption{\label{fig:NjV} The number of jets with minimum $p_{T}~>~$ 20~GeV assigned to the \Vframe~frame, \NjV, after application of the $\PTISR$ and $\RISR$ selections described in Table~\ref{Tab:SignalRegions}. Gluino signals tends to have a larger \NjV compared to SM backgrounds.}
\end{figure}

As is typical in searches for squarks and gluinos, selection requirements based on reconstructed jet multiplicity can suppress contributions from backgrounds with characteristically fewer jets, such as di-boson and vector-boson + jets processes. Using the decay tree interpretation imposed on each event, the efficacy of such requirements can be enhanced by taking into account the partitioning of jets between the \Vframe and \ISRframe systems. While the multiplicity of \ISRframe-associated jets tends to be similar between signals and backgrounds, the number of jets in each event assigned to the \Vframe system, \NjV, is a powerful discriminant, as demonstrated in Figure~\ref{fig:NjV}. Increasing mass-splittings between parent and daughter sparticles result in, on average, larger \NjV, with cuts on this observable suppressing vector boson + jets backgrounds in particular.

\begin{figure}[htb!]
\centering 
\includegraphics[width=.235\textwidth]{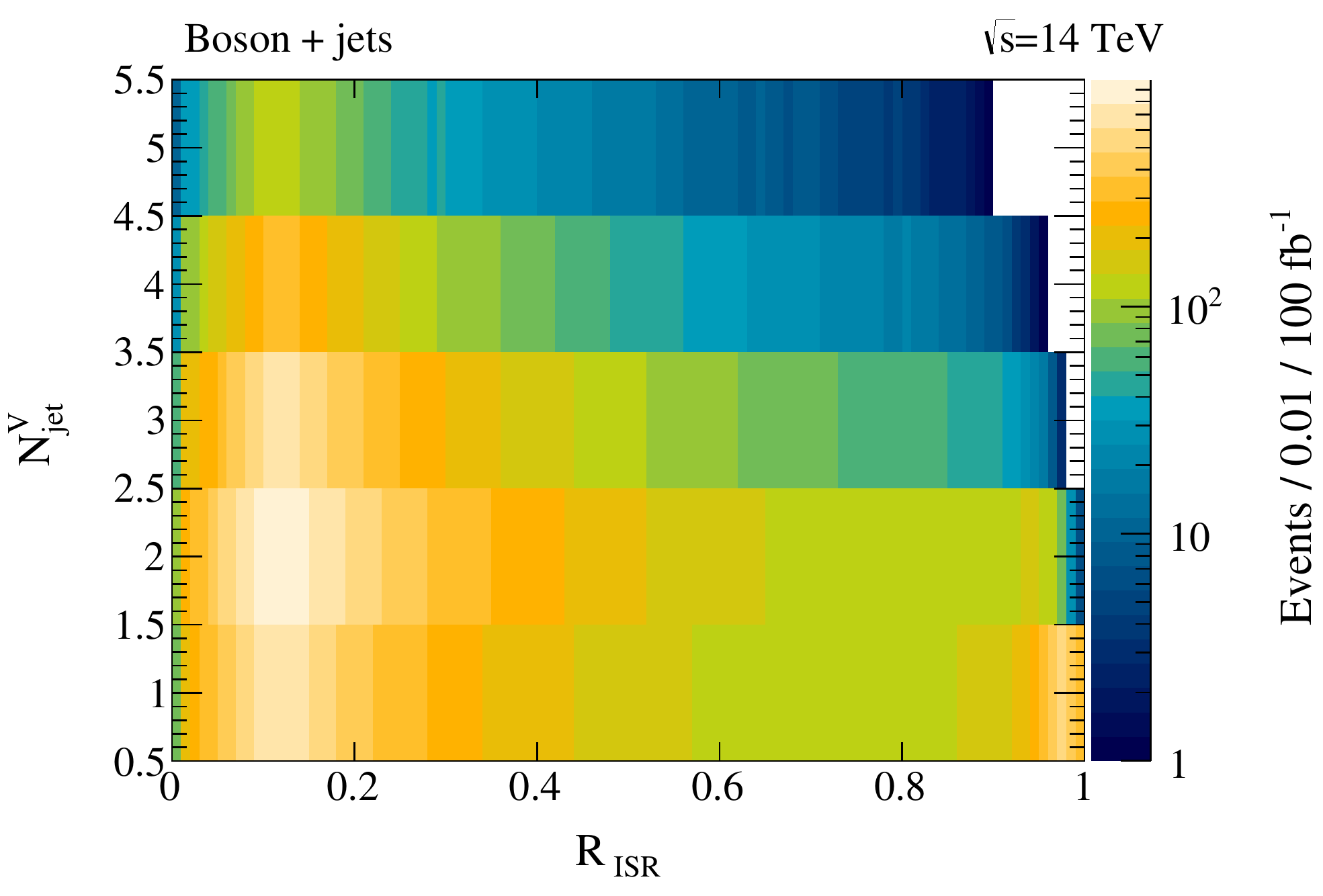}
\includegraphics[width=.235\textwidth]{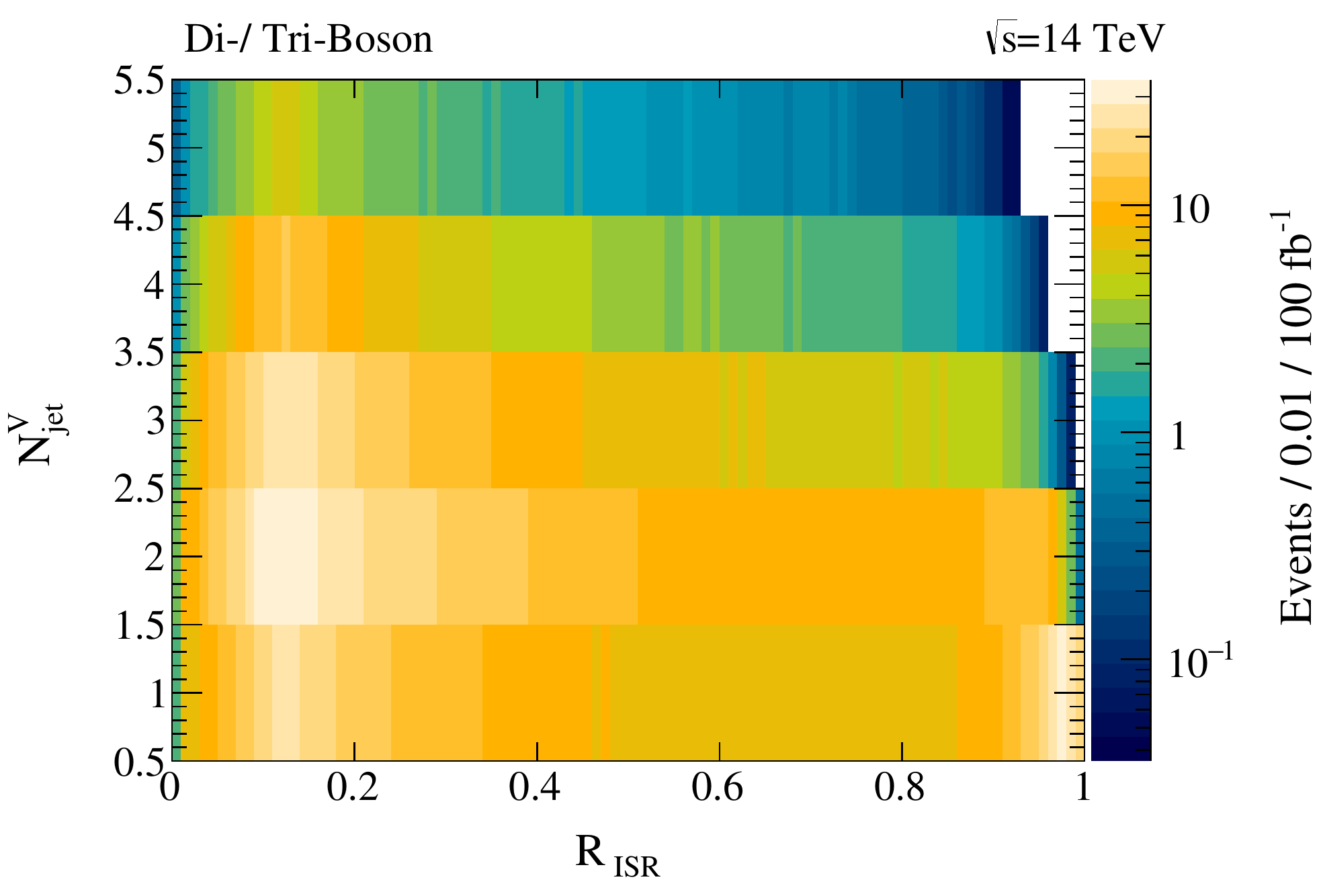}
\includegraphics[width=.235\textwidth]{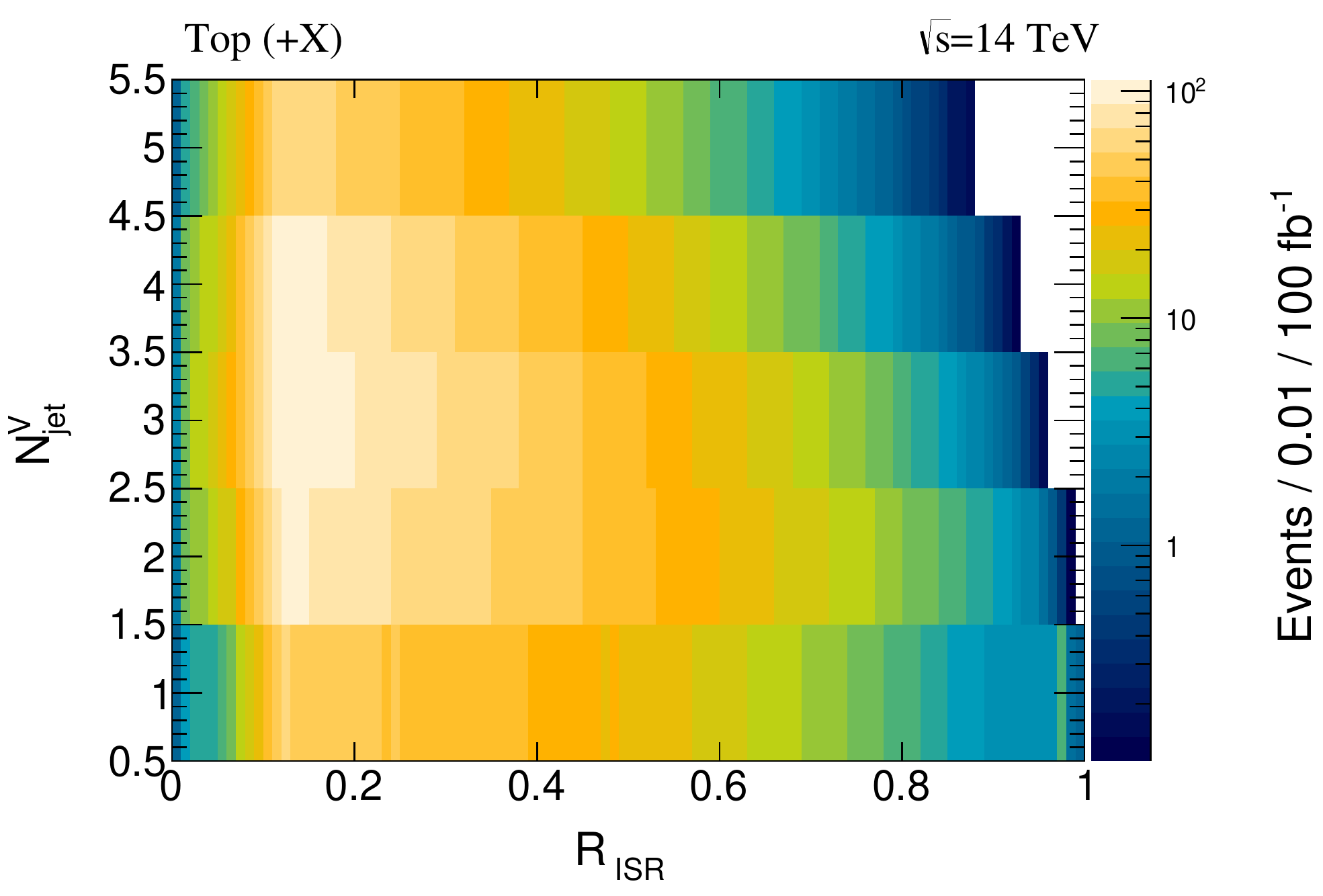}
\includegraphics[width=.235\textwidth]{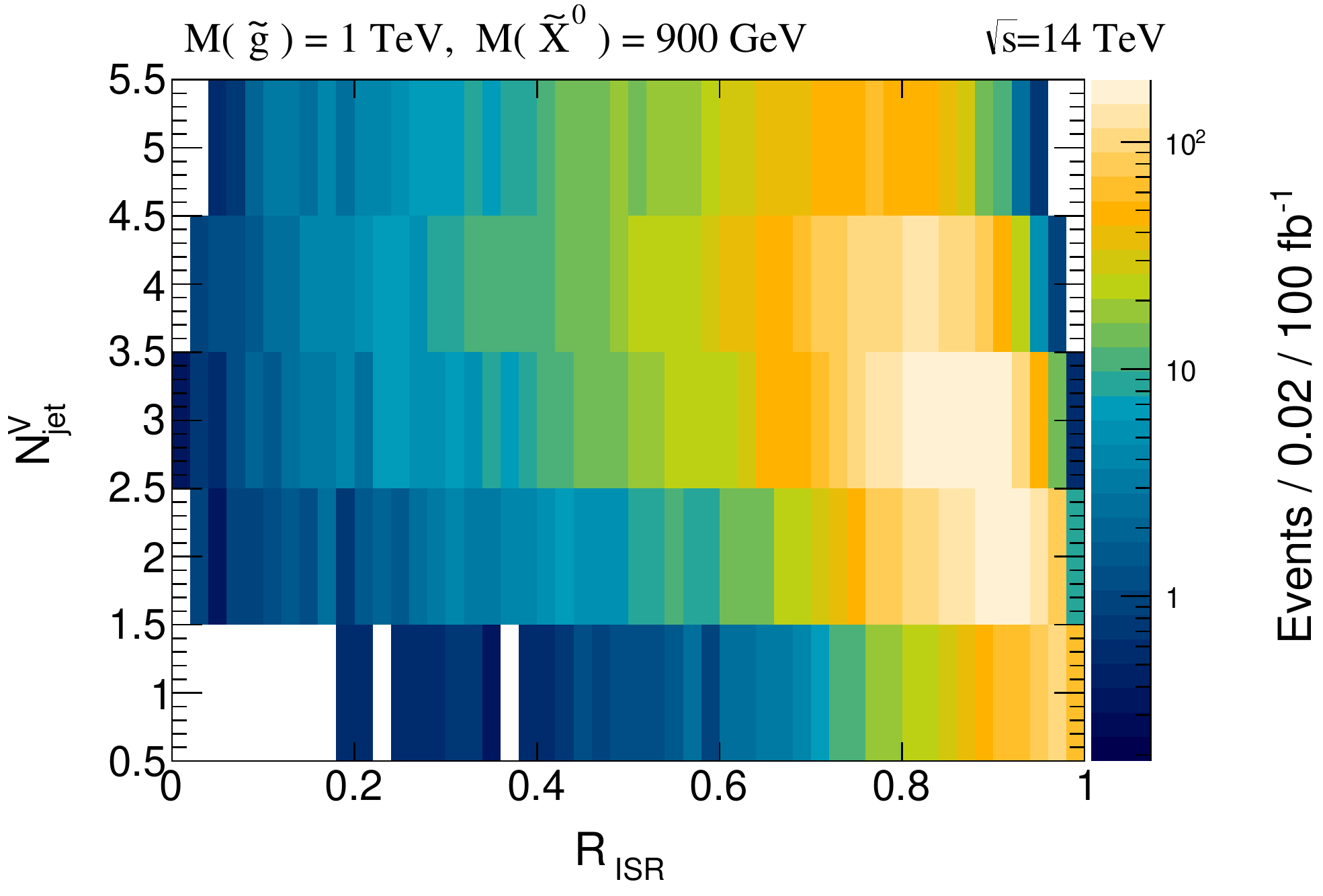}
\caption{\label{fig:RISR_vs_NjV}  
Distribution of Number of jets with momentum greater than 20~GeV, assigned to the Visible system `\Vframe' as a function of \RISR  for the boson+jets (upper left), di-boson (upper right), top+X (lower left) and gluino signal (lower right) samples.}
\end{figure}

The complementarity of the \NjV selection requirement with \RISR is illustrated in Figure~\ref{fig:RISR_vs_NjV}, where the two-dimensional distribution of \NjV~and \RISR~is plotted for vector boson+jets, di-boson and top+X, to be compared with a gluino signal
with a mass-splitting of 100~GeV. Notably, the large background contributions from boson+jets and di-boson in the high \RISR~region occur at $\NjV=1$. These processes include $W\rightarrow\tau\nu$, where the hadronically decaying $\tau$ lepton is mis-identified as a jet, and $Z(\rightarrow \nu\bar{\nu}$) events with only one associated jet in the \Vframe~system.
For gluino signals, a minimum of three jets associated with the sparticle system is a favorable selection, whereas cases where the mass-splitting is larger tend to benefit from the imposition of an even tighter requirement.  
A useful anti-correlation between \RISR and \NjV can be exploited to define different signal regions, benefitting from their interplay.

\begin{figure}[htb!]
\centering 
\includegraphics[width=.235\textwidth]{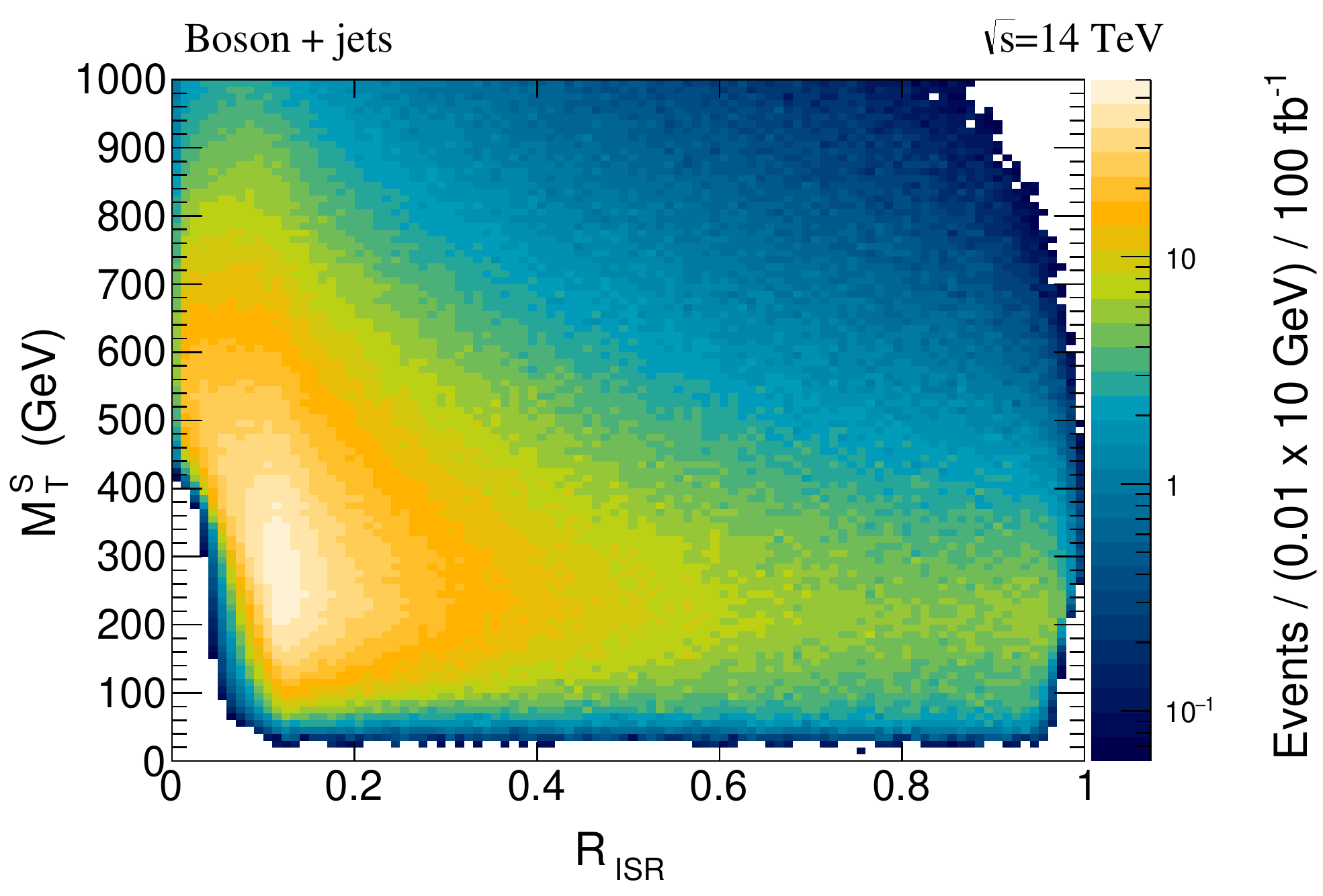}
\includegraphics[width=.235\textwidth]{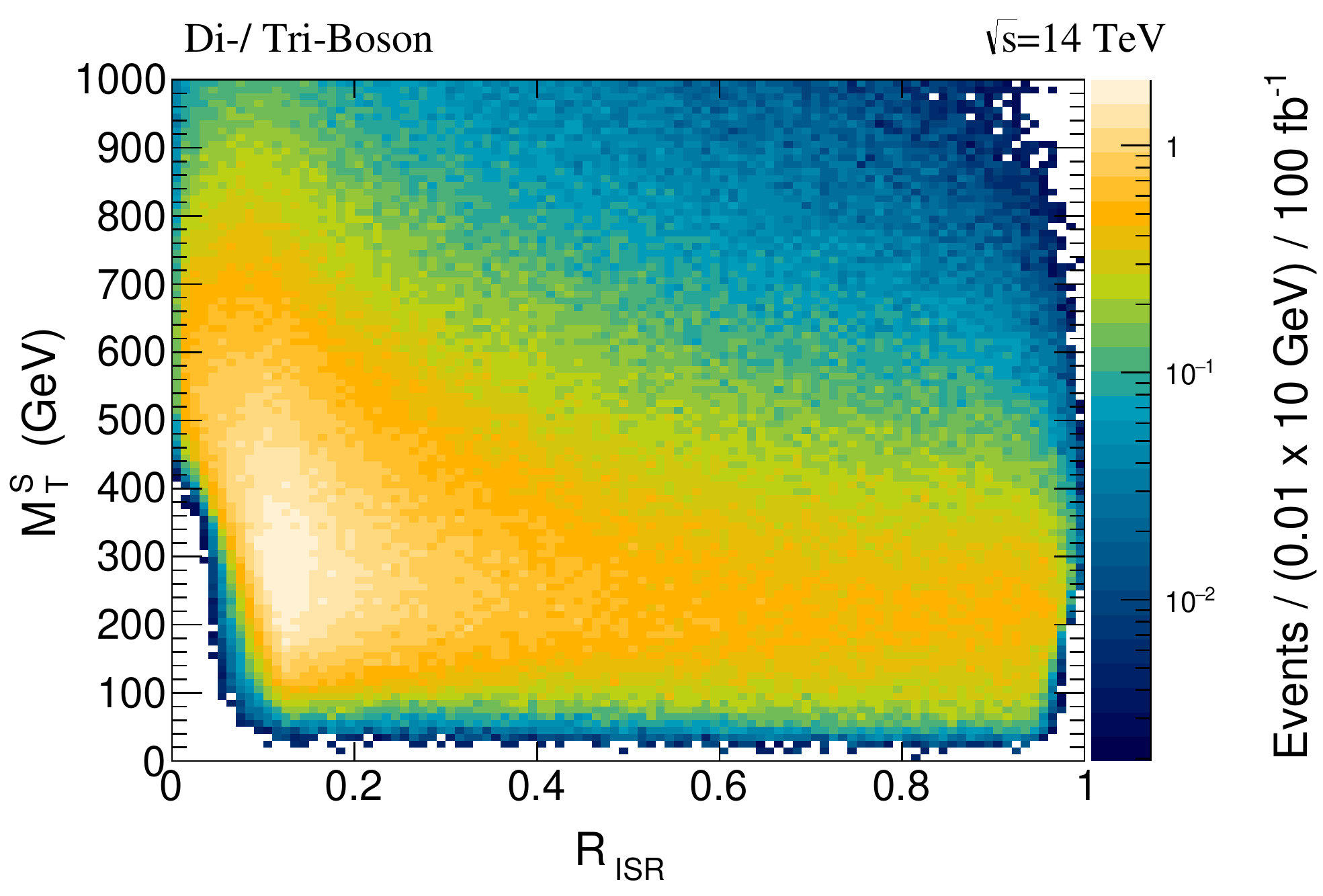}
\includegraphics[width=.235\textwidth]{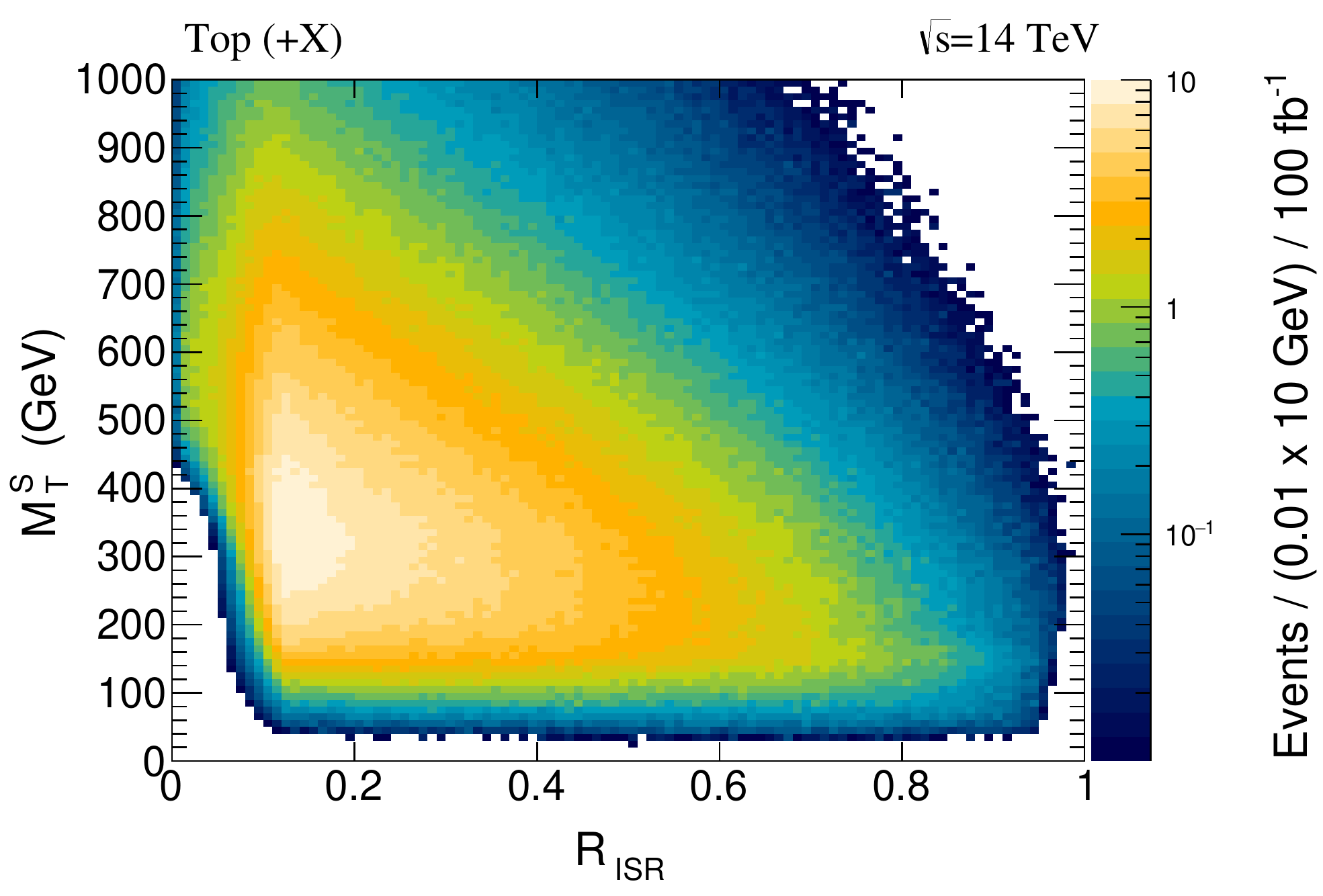}
\includegraphics[width=.235\textwidth]{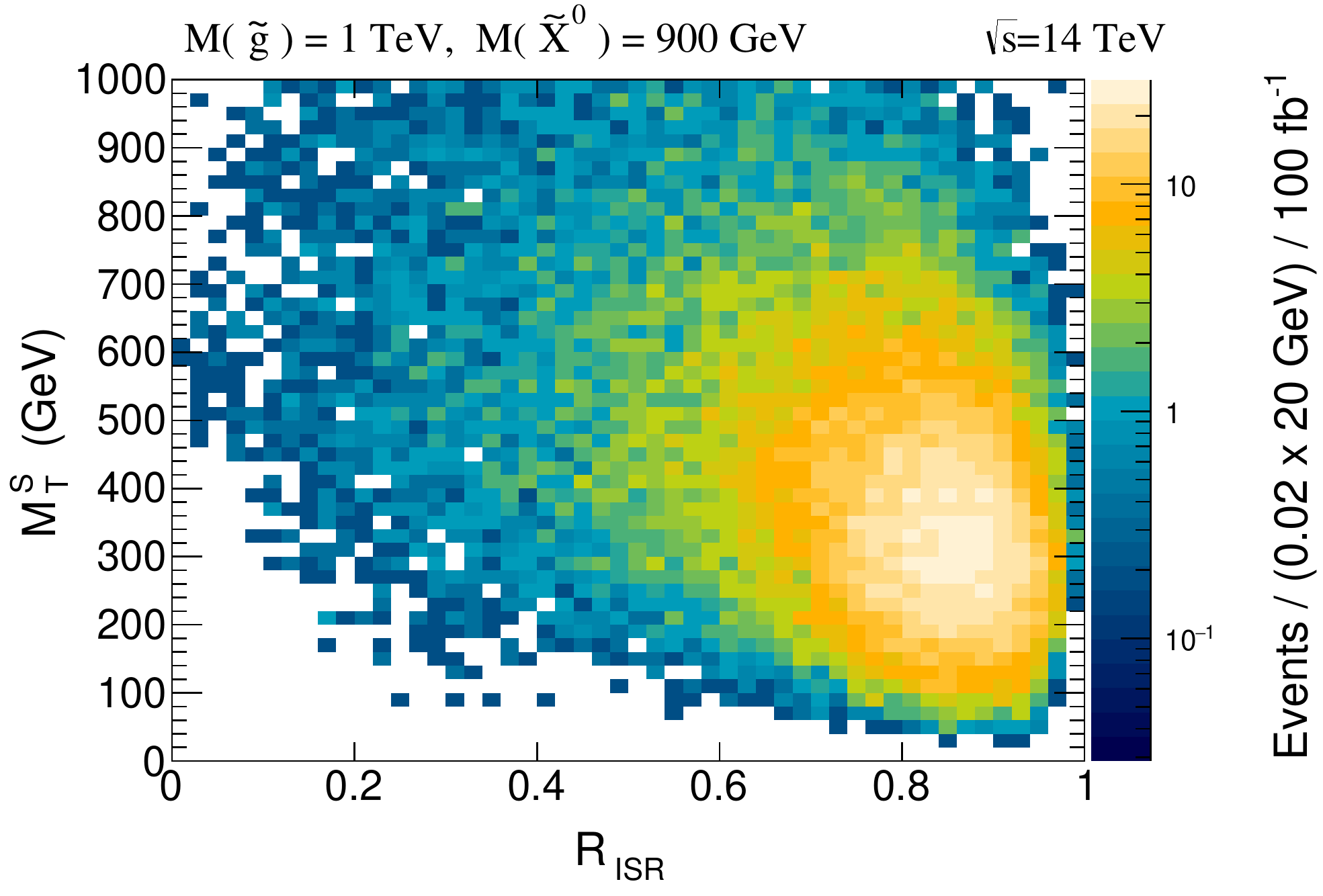}
\caption{\label{fig:RISR_vs_MS}  
Distribution of transverse mass of the `sparticle' system `\Sframe' (\MS) as a function of \RISR for the boson+jets (upper left), di-boson (upper right), top+X (lower left) and gluino signal (lower right) samples.}
\end{figure}

The distribution of the transverse mass of all the constituents of the \Sframe system, \MS, as a function of \RISR, can be seen in Figure~\ref{fig:RISR_vs_MS}, after imposing the requirements on \PTISR and \NjV from Table~\ref{Tab:SignalRegions}.
While the discrimination power of \RISR is visible, it is clear that background contributions can be successfully managed by relaxing the selection on \RISR in favor of an additional requirement on \MS. Tightening the selection requirement on \MS with increasing signal mass-splittings can be used to compensate for decreasing \RISR discrimination.

\begin{figure}[htb!]
\centering 
\includegraphics[width=.46\textwidth]{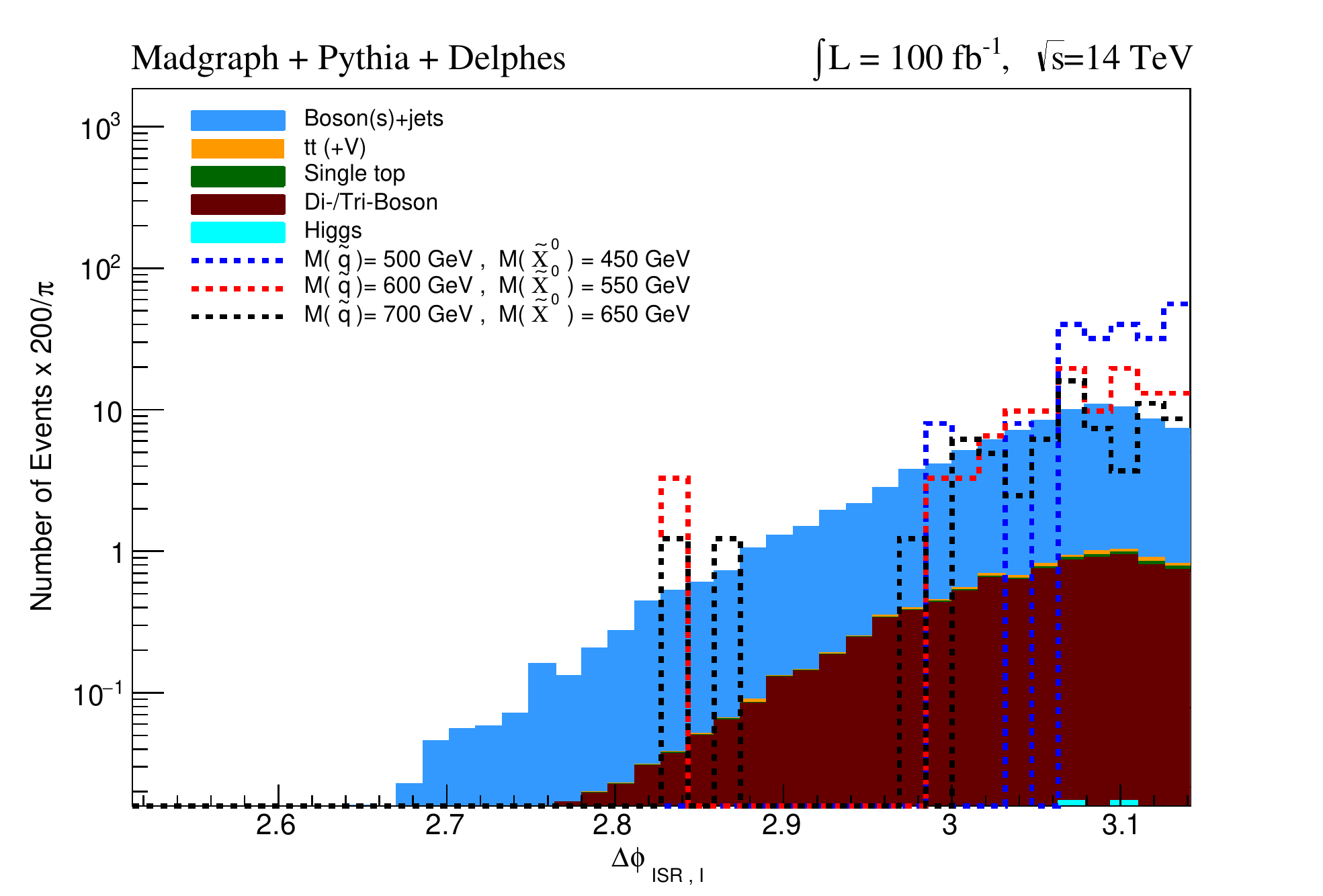}
\caption{\label{fig:dphiISRI} Distribution of \dphiISRI for SM backgrounds compared with the squark pair signal samples. All curves are shown after applying the relevant criteria from Table~\ref{Tab:SignalRegions}. Although both signal and background distributions tend towards $\pi$ the signal has a much stronger tendency to do so.}
\end{figure}

The final variable considered from the application of Recursive Jigsaw Reconstruction to the decay tree in Figure~\ref{fig:decayTree} is the opening angle between the \ISRframe~system and the invisible system evaluated in the \CMframe~frame, \dphiISRI. The distribution of \dphiISRI, after the application of the loosest of all other selection criteria on other variables in Table~\ref{Tab:SignalRegions}, is shown for example squark signals and backgrounds in Figure~\ref{fig:dphiISRI}.
Further improvements in sensitivity can be achieved through a requirement on this observable. In order to maintain a conservative selection, the same requirement of \dphiISRI~$>$~3 is imposed for gluinos and squarks in all signal scenarios studied. To extract the optimal significance for each signal point considered, one could consider further tuning the selection criteria on this quantity. 

\begin{figure}[htb!]
\centering 
\includegraphics[width=.46\textwidth]{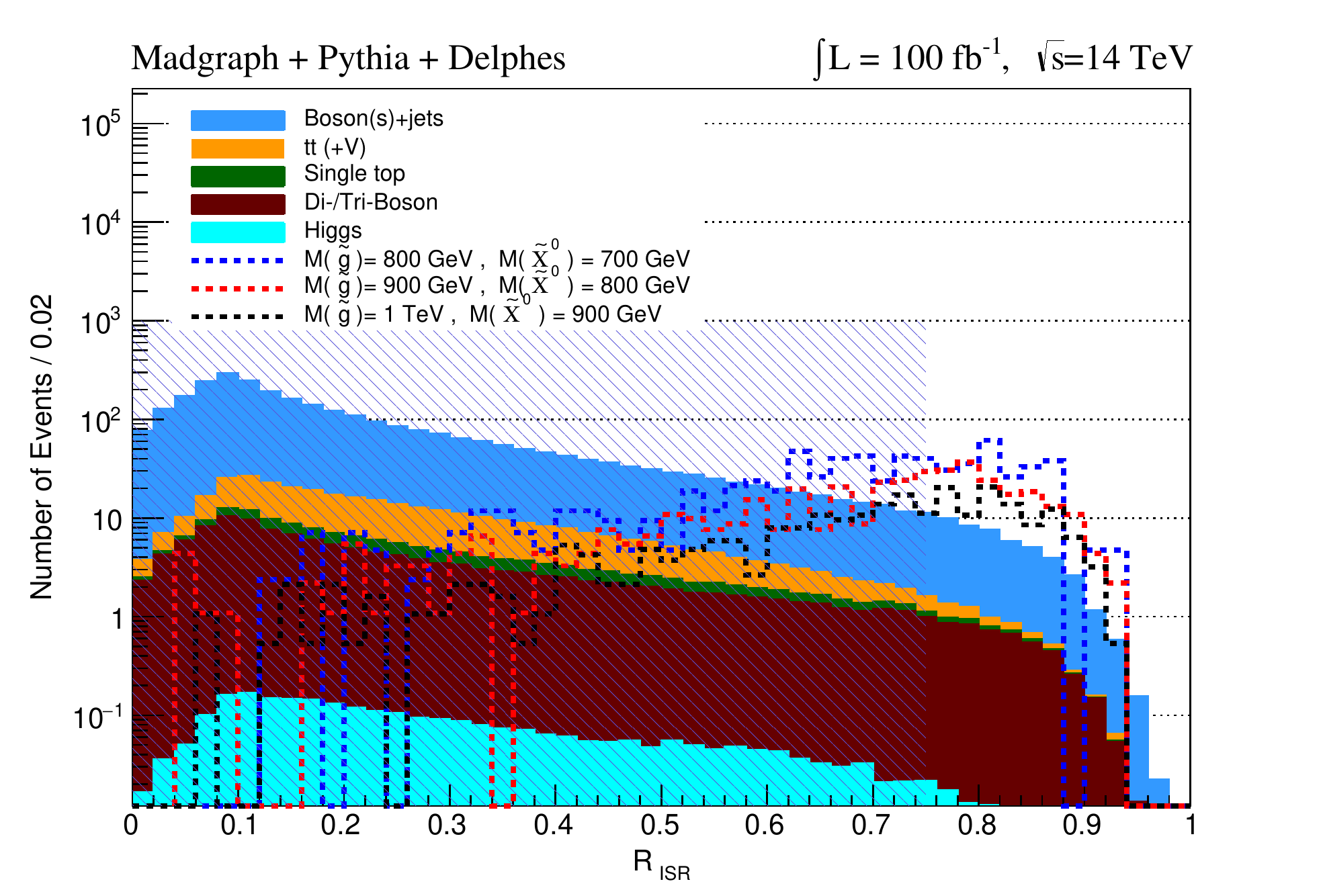}
\caption{\label{fig:SigRegion} The distribution of \RISR for example gluino signals and backgrounds after the application of requirements on \PTISR ($>$ 1000~GeV),  \MS ($>$ 100~GeV), \NjV $\geq 3$, and $\dphiISRI (>$~3.0). We observe that selecting events with large values of \RISR provides excellent discrimination between signals and backgrounds.}
\end{figure}

Taking into account the phenomenology of the variables considered, and studying sensitivity variations with different selection requirements, we define four example signal regions aimed at targeting the different mass-splittings scenarios represented by the simulated signal samples. We summarize the `preselection criteria' in one line, including a veto on events containing an electron, muon or jet tagged as having been initiated by the fragmentation of a $b$-quark, along with loose requirements on the \met~($>$100~GeV)
and leading jet transverse momentum ($p_{T} >~20$ GeV). Events where large momentum is provided to the sparticles from ISR are selected by considering only those with \PTISR~$>$~1000~GeV. In this high \PTISR regime there is an interesting interplay between the further selection criteria considered. The \RISR~selection applied is progressively looser as the target signal mass-splitting becomes larger, with correspondingly more stringent requirements on the \MS~and \NjV~variables. 
A conservative \dphiISRI requirement is applied to all events. Furthermore, we apply a minimum transverse momentum requirement on the second (third) jet in the squark (gluino) analysis. There are modest but unique gains from selecting events
with increasingly larger values of the momenta of these jets as a function of increased mass-splitting between parent sparticles and LSPs. Representative signal and background yields for 100 fb$^{-1}$ of data at the 14 TeV LHC after the application of these requirements are illustrated in Figure~\ref{fig:SigRegion}.

\newpage

\onecolumngrid
\begin{center}
\begin{table}[htb!]
{\small
\begin{tabular}{|c|c|c|c|c|}
\hline
$\downarrow$Variable $\setminus$ Mass Splitting$\rightarrow$ & $\Delta m=25$ GeV & $\Delta m=50$ GeV & $\Delta m=100$ GeV & $\Delta m=200$ GeV\tabularnewline
\hline
\hline
Preselection criteria & \multicolumn{4}{c|}{lepton ($e$ and $\mu$) and $b$-jet veto, $\met~>~100$ GeV, $p_{T}^{~\mathrm{jet}} >~20$ GeV}\tabularnewline
\hline
\PTISR~{[}GeV{]} & \multicolumn{4}{c|}{$>1000$}\tabularnewline
\hline
\RISR & $>0.9$ & $>0.85$ & $>0.75$ & $>0.65$\tabularnewline
\hline
\MS~{[}GeV{]} & $-$ & $>100$ & $>250$ & $>400$\tabularnewline
\hline
\NjV & \multicolumn{2}{c|}{$\geq3$ ($\geq2$)} & \multicolumn{2}{c|}{$\geq4$ ($\geq2$)}\tabularnewline
\hline
\multicolumn{1}{|c|}{ $p_{T}^{~\mathrm{jet}3,V}$ $(p_{T}^{~\mathrm{jet}2,V})$~{[}GeV{]}} & $>20$ ($>40$) & $>30$ ($>60$) & $>40$ ($>120$) & $>50$ ($>140$)\tabularnewline
\hline
\dphiISRI  & \multicolumn{4}{c|}{$>3.0$}\tabularnewline
\hline
\end{tabular}}
\caption{A conservatively optimized set of selection criteria for signal regions in the analysis of gluino (squark) pair-production. The selection assumes a sample of 100 fb$^{-1}$ collected in proton-proton collisions with a center-of-mass energy of $\sqrt{s}=14$~TeV. The natural pattern is to loosen the scaleless criteria as the criteria with units GeV are tightened. \label{Tab:SignalRegions}}
\end{table}
\end{center}
\twocolumngrid

\section{Results and Summary \label{sec:summary}}

\begin{figure}[htb!]
\centering 
\includegraphics[width=.37\textwidth]{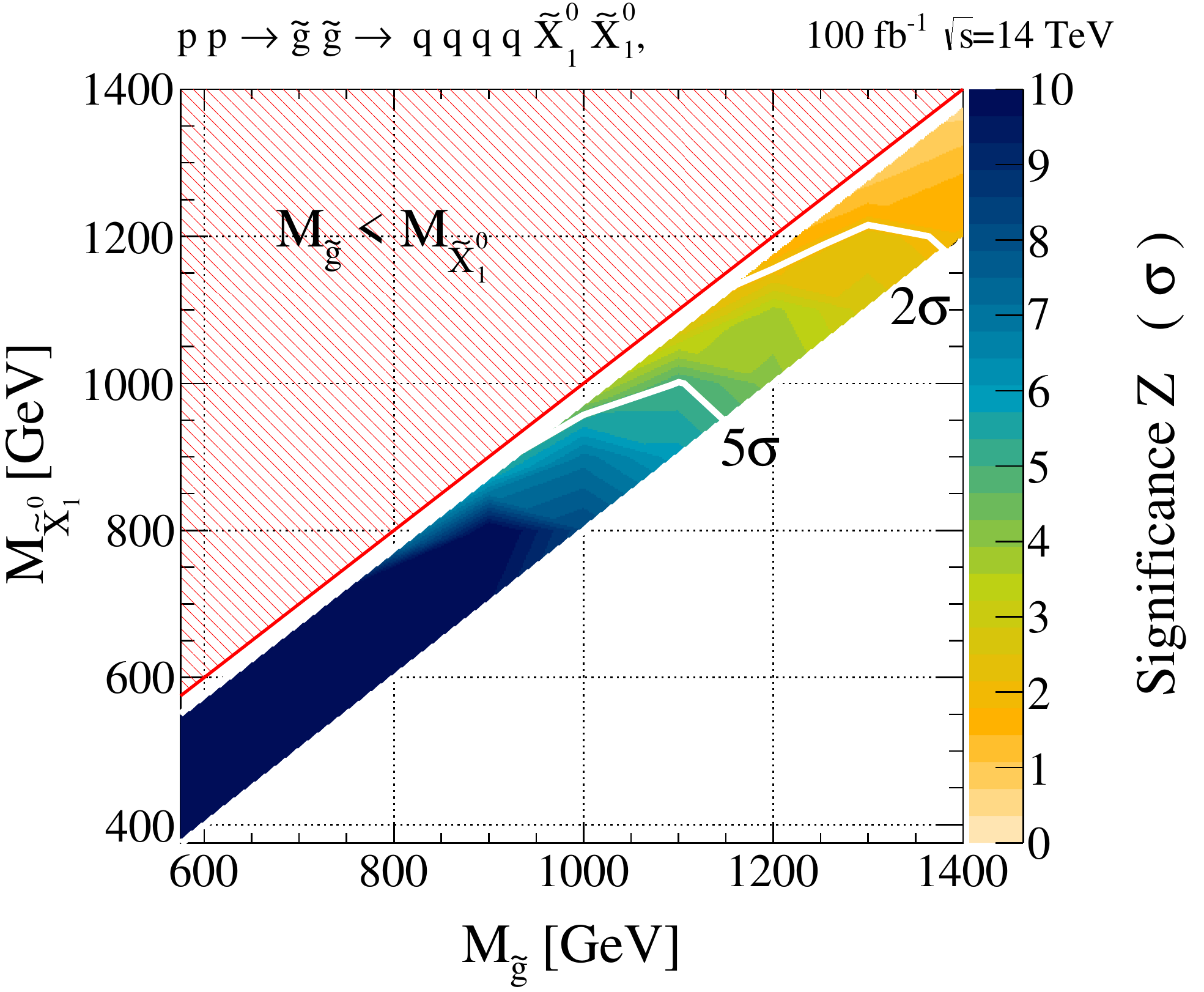}
\includegraphics[width=.37\textwidth]{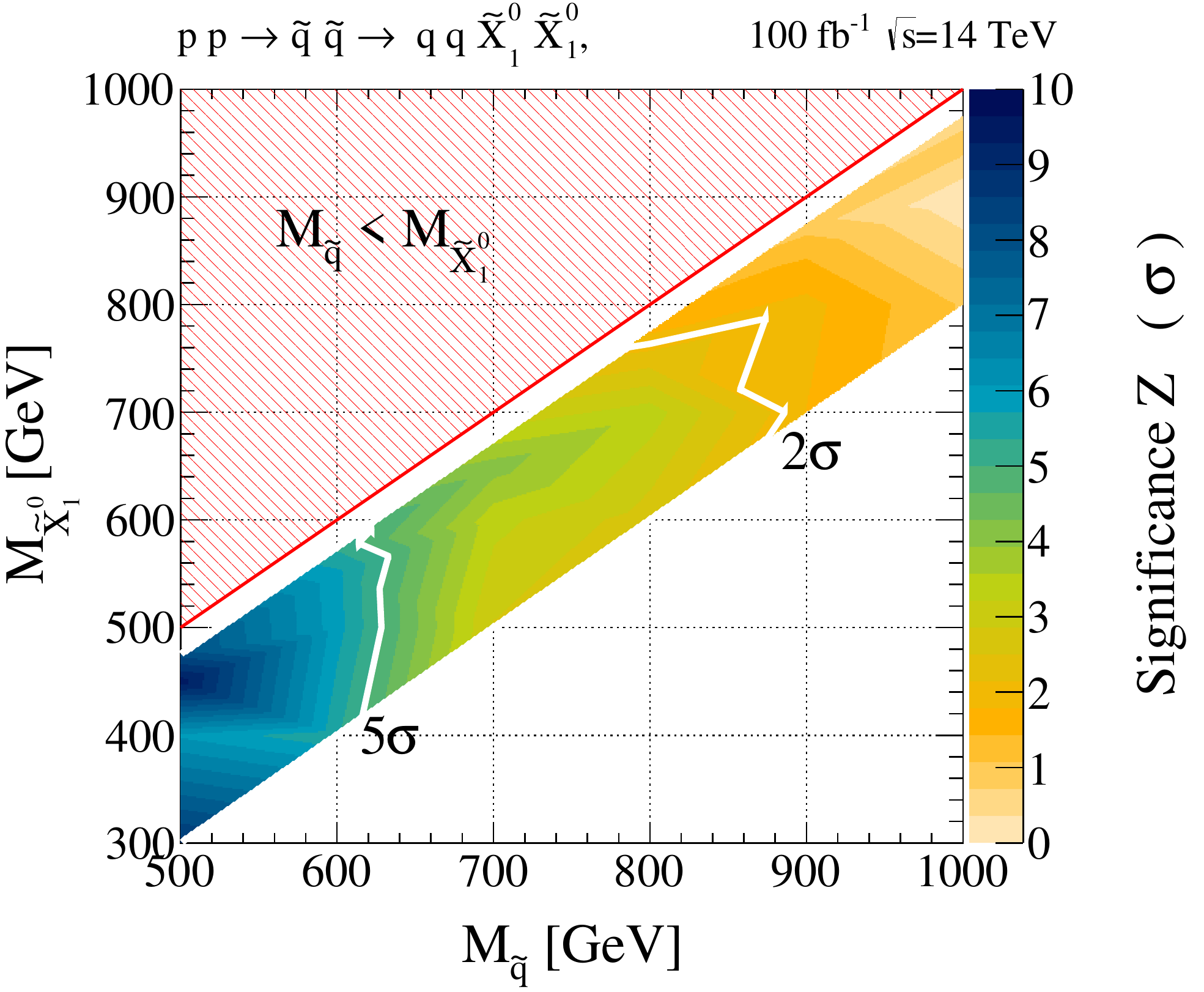}
\caption{\label{fig:significance} Projected exclusion and discovery reach for gluino pair-production (upper plot) squark pair-production (lower plot) in the compressed regions with $25$~GeV $\leq\Delta m \leq 200$~GeV.}
\vspace{-0.4cm}
\end{figure}

Applying the selection criteria from Table~\ref{Tab:SignalRegions}, we calculate projected sensitivities for putative gluino and squark signals with compressed mass spectra, shown in Figure~\ref{fig:significance}. 
The significance of a given signal in the presence of background is calculated using $Z_{B_{i}}$ as a metric~\cite{ref:Zbi}. The required inputs are the signal and background yield, statistical uncertainties and an assumed background systematic uncertainty of 15$\%$ for all scenarios considered.
With an integrated luminosity of 100 fb$^{-1}$, corresponding to proton-proton collisions with a center-of-mass energy of $\sqrt{s}=14$~TeV, gluinos in compressed scenarios with masses above 1~TeV can be discovered, with exclusion significance for masses of 1.4~TeV in some cases. Squarks with masses of 600~GeV and mass-splittings up to 200~GeV would be discovered greater than 5$\sigma$ significance, while they can be excluded for masses between $\sim$800~GeV and 900~GeV.

We have introduced a new approach to searches for compressed SUSY signals based on the imposition of a simple decay tree interpretation on reconstructed events, using the Recursive Jigsaw Reconstruction technique~\cite{ref:basis}. Herein we have focused on signals of di-squark and di-gluino production with final states rich in jets and weakly interacting sparticles. By leveraging 
a high momentum system of initial state radiation to boost potential sparticles in events, we can construct a basis of experimental observables that are sensitive to the mass-splittings between the parent and child sparticles and that are capable
of discriminating against SM background in otherwise challenging final states. In particular, we demonstrate that the transverse momentum of ISR systems works in concert with the \RISR~variable to limit the yields of background processes 
and provide unique additional information.  
Although the primary focus of this work has been on solely hadronic final states, we also demonstrate that the approach can trivially be applied to {\it any final state}.
By simply assigning to the $\Sframe$~system the reconstructed particles consistent with the expected decays of sparticles, with additional jets associated to the \ISRframe~system using the provided prescription, 
analyses targeting compressed final states with $b$-jets, charged leptons, photons, and other objects can benefit from the same methods. For each of these cases, the spirit of the approach remains unchanged and the additional handles of object identification and multiplicity can prove valuable in reducing backgrounds still further.

The code to reproduce this decay tree (see Figure~\ref{fig:decayTree}), set up the framework, and calculate the observables of interest is made available through the \texttt{RestFrames} package~\cite{ref:restframes}.

\begin{acknowledgments}
The authors would like to thank the Harvard Club of Australia for the award of the ``Australia-Harvard Fellowship'' in 2014 and 2015 which 
made part of this work possible. PJ is supported by the Australian Research Council Future Fellowship FT130100018.
CR wishes to thank the Harvard Society of Fellows and William F. Milton Fund for their generous support.
\end{acknowledgments}


\begin{thebibliography}{99}

\bibitem{ref:Golfand1971iw} Y.~A. Golfand and E.~P. Likhtman, JETP Lett. {13} (1971)  323.
\bibitem{ref:Neveu1971rx} A.~Neveu and J.~H. Schwarz, \href{http://dx.doi.org/10.1016/0550-3213(71)90448-2}{Nucl. Phys. {B31} (1971)  86}.
\bibitem{ref:Ramond1971gb} P.~Ramond, \href{http://dx.doi.org/10.1103/PhysRevD.3.2415}{Phys. Rev. {D3} (1971)  2415}.
\bibitem{ref:Volkov1973ix} D.~V. Volkov and V.~P. Akulov, \href{http://dx.doi.org/10.1016/0370-2693(73)90490-5}{Phys. Lett. {B46} (1973)  109}.
\bibitem{ref:Wess1974tw} J.~Wess and B.~Zumino, \href{http://dx.doi.org/10.1016/0550-3213(74)90355-1}{Nucl. Phys. {B70} (1974) 39}.

\bibitem{ref:compressed_stop-1} H. An and L-T. Wang, Phys. Rev. Lett. {\bf 115}, 181602 (2015).
\bibitem{ref:compressed_stop-2} S. Macaluso, M. Park, D. Shih and B. Tweedie, JHEP 1603 (2016) 151.
\bibitem{ref:compressedgluinos} A.~Delgado, A.~Martin and N.~Raj, arXiV:1605.06479.

\bibitem{ref:basis} P.~Jackson, C.~Rogan, \emph{The Recursive Jigsaw Reconstruction} - In preparation.

\bibitem{ref:restframes} C.~Rogan, \emph{RestFrames}, http://RestFrames.com

\bibitem{ref:snowmass-1} J.~Anderson {\it et. al.} \emph{Snowmass Energy Frontier Simulations}, FERMILAB-TM-2566-CMS-E-PPD.
\bibitem{ref:madgraph} J.~Alwall, R.~Frederix, S.~Frixione et al. JHEP 1407 (2014) 079.
\bibitem{ref:pythia} T. Sjostrand, S. Mrenna, and P. Skands, JHEP 0605 (2006) 026.
\bibitem{ref:delphes} J.~ de Favereau et al., JHEP 1402 (2014) 057.
\bibitem{ref:atlas} ATLAS Collaboration, JINST 03 S08003 (2008).
\bibitem{ref:cms} CMS Collaboration, JINST 03 S08004 (2008).
\bibitem{ref:snowmass-2} A. Avetisyan et al., arXiv:1308.1636; A. Avetisyan et al., arXiv:1308.0843.
\bibitem{ref:Zbi} R.~Cousins et al., NIMPA A595 (2008) 480-501.

 



\end{thebibliography}

\end{document}